\documentclass[runningheads]{llncs}
\usepackage{graphicx}
\usepackage{amsmath}
\usepackage{amssymb}
\usepackage{color}

\newcommand{\hide}[1]{ }

\usepackage{graphicx}
\usepackage{subfigure}
\usepackage{amsmath}
\usepackage{color}
\usepackage{url}
\usepackage{algorithm}  
\usepackage{algorithmic} 
\usepackage{multirow}

\begin{document}
\title{Learning Safe Neural Network Controllers \\ with Barrier Certificates
}
%
%
\author{Hengjun Zhao\inst{1} \and
Xia Zeng\inst{1} \and
Taolue Chen\inst{2}\and
Zhiming Liu\inst{1}\and
Jim Woodcock\inst{1,3}}
\authorrunning{H. Zhao et al.}
%
\institute{
School of Computer and Information Science, Southwest University\\
\email{\{zhaohj2016, xzeng0712, zhimingliu88\}@swu.edu.cn}
\and
Department of Computer Science, University of Surrey\\
\email{taolue.chen@surrey.ac.uk}
\and
Department of Computer Science, University of York\\
\email{jim.woodcock@york.ac.uk}
}
\maketitle              

\begin{abstract}
We provide a novel approach to synthesize controllers for nonlinear continuous dynamical systems with
control against safety properties. The controllers are based on neural networks (NNs).
To certify the safety property we utilize barrier functions, which are  represented by NNs as well.
We train the controller-NN and barrier-NN simultaneously, achieving a verification-in-the-loop synthesis.
We provide a prototype tool {\sf nncontroller} with a number of case studies. 
The experiment results confirm the feasibility and efficacy of our approach.
\end{abstract}


\section{Introduction}
Controller design and synthesis is one of the most fundamental problems in control theory. In recent years,
especially with the boom of deep learning, there has been considerable research activities in the use of
neural networks (NNs) for control of nonlinear systems \cite{DeepM-DRL,OpenAI-DRL}. NNs
feature the versatile representational ability of nonlinear maps and fast computation, 
making them an ideal candidate for sophisticated control tasks \cite{NNforControl}.
Typical examples include self-driving cars, drones, and
smart cities. It is noteworthy that many of these applications 
are safety-critical systems, where safety refers to, in a basic form, that
the system cannot reach a dangerous or unwanted state.
For control systems in a multitude of Cyber-Physical-System domains,
designing \emph{safe} controllers which can guarantee safety behaviors of
the controlled systems is of paramount importance
\cite{saferl-openai,Berkenkamp17-nips,lya-nn-2018,DUTTA2018151,taylor2019learning,End-to-end19,choi2020reinforcement,yaghoubi2020training,PappasHSCC20,NNV-CAV20}.

Typically, when a controller is given, formal verification is required to certify its safety.
Our previous work \cite{ZhaoZC020} has dealt with the \emph{verification} of continuous dynamical systems
by the aid of neural networks. In a nutshell, we follow a deductive verification methodology therein
by synthesizing a barrier function, the existence of which suffices to show the safety of the controlled
dynamical system. The crux was to use neural networks to represent the barrier functions, spurred by  the well-known universal approximation theorem \cite{LESHNO1993861} which assures the expressibility of NNs.   

It is imperative to realize that verification or certification of an existing controller does not lend itself to effective and efficient \emph{construction} of controllers, which is the main focus of the current work. Following a correctness-by-design methodology, we aim to synthesize controllers
which can guarantee that the controlled system is safe. This question is considerably more challenging and perhaps
more interesting from a system engineering perspective. To this end we adopt a data-driven approach for the design of controllers which are to be represented as an NN. 
A key issue of controller synthesis is to provide a formal guarantee of the quality for the obtained controller, of which safety is arguably the most fundamental. A common practice is to first come up with a controller and then to verify it against desired properties. An interesting innovation of our work is, however, to integrate the synthesis and verification in a unified, data-driven framework, which is enabled by our earlier work by using NNs as a certification mechanism. At a high level, our approach for the controller synthesis will produce  two neural networks simultaneously, i.e., 
one is used to represent the controller (henceforth referred to as controller-NN),
and the other is used to represent the barrier function (henceforth referred to as barrier-NN).
The synergy of the two NNs, supported by an additional verification procedure to make sure the learned barrier-NN is
indeed a barrier certificate, provides the desired safety guarantee for the synthesized controller.  

Our method follows a data-driven framework in the sense  that both NNs are trained from datasets. For that purpose, we generate training sets and propose  specifically designed  loss functions which are the key towards application of standard learning algorithms for NNs.  
%
In terms of the learned NN controllers, we find that they usually respect safety constraints, but may exhibit poor performance in terms of, 
e.g., stability. To further improve the synthesized controllers, we propose a number of approaches such as imposing a larger safety region, stability-aware loss functions, and bounded control inputs (via the Hardtanh activation function). 

In general, the advantages of our approach are threefold: (1) the approach is data-driven, requiring considerably less control
theory expertise; (2) the approach can support non-linear control systems and safety properties,
owing to the representation power of neural networks; and (3) the approach can achieve verification-in-the-loop synthesis,
owing to the co-synthesis of controller and barrier functions, which can be seamlessly integrated to provide a
correctness-by-design controller as well as its certification.

The main contributions of the   paper are summarized as follows:
\begin{itemize}
	\item We put forward a learning-based framework to synthesize controllers as well as the associated safety certification. This is largely a data-driven approach, with little prior knowledge required, and enjoys great flexibility to effectively handle nonlinear (beyond polynomial) dynamics of ODEs. 
	
	%
	\item We instantiate the framework by using new class of activation functions. 
  Moreover, we demonstrate how to generate training set, and to construct loss functions of neural networks. 
  We also  provide practical methods to formally verify  the learnt barrier certificates represented as neural networks. 
	%
	\item We carry out proof-of-concept case studies to showcase the efficacy of the approach. 
\end{itemize}
%

\subsection{Related Work}
Our work on learning and verifying NN controllers with barrier certificates is closely related to two
categories of research, i.e. \emph{safety critical control by machine learning} and \emph{formal verification of neural networks}.
Note that the discussions below are necessarily non-exhaustive as a reasonably detailed discussion requires an independent survey.  

\paragraph{Safety Critical Control by Machine Learning.} Research work in this category has been emerging in the past  years.
They differ in: (1) the overall learning framework, e.g. reinforcement learning (RL) or supervised learning;
(2) the component to be learned (especially by NN), e.g., the system model, the feedback control policy, or the safety certificate; (3) the kind of safety
certificate, e.g., control Lyapunov function (CLF)  or control barrier function (CBF) \cite{CBF19}. 
A verification-in-the-loop RL algorithm was proposed in \cite{Deshmukh2019LearningDN} to learn safe NN controllers for known system dynamics using CBFs;
an end-to-end safe RL architecture was developed by combining model-free RL control, model-based CBF control, 
and model learning in \cite{End-to-end19}; CLFs and CBFs are integrated into the episodic learning framework and RL framework with 
an emphasis on  model uncertainties in \cite{Episodic19,taylor2019learning,choi2020reinforcement}; CBFs are integrated with imitation learning to train 
safe NN controllers in \cite{yaghoubi2020training}. For all the above work, CLFs or CBFs are assumed to be given, at least in a parametric form.
For CLFs or CBFs synthesis, a demonstrator-learner-verifier framework was proposed in \cite{lya-sriram-19} to learn polynomial CLFs for polynomial nonlinear dynamical systems;
a special type of neural network was designed in \cite{lya-nn-2018} as candidates for learning Lyapunov functions; a supervised learning approach was proposed in
\cite{GAO-NIPS2019} to learn neural network Lyapunov functions and linear control policies; data-driven model predictive control (MPC) exploiting neural 
Lyapunov function and neural network dynamics model was proposed in \cite{DUTTA2018151,mittal2020neural}. For multi-agent systems, 
barrier function has recently been applied for safe policy synthesis on POMDP models \cite{ahmadi2019safe}. The computer science community
has dealt with the issue of safe controller learning in different ways from above: for example, a logical-proof based approach was proposed in \cite{FultonP18}
towards safe RL; a synthesis framework capable of synthesizing deterministic programs from neural network policies was proposed in \cite{PLDI2019HZhu} and so 
formal verification techniques for traditional software systems can be applied. 
Compared with these works, our approach has the following features which make it unique:
\begin{itemize}
    \item controller and safety certificate are both represented and learned by NNs of general structure; 
            no prior knowledge or initial guess is required;
    \item training data generation is based on state space sampling, and therefore trajectory simulation is not needed;
    \item although the method is based on known dynamics, we believe it is possible to extend it with dynamics learning by introducing the third NN representing
        system dynamics \cite{manek2020learning}. 
\end{itemize}

\paragraph{Formal Verification of Neural Networks.}
This has attracted considerable research efforts in recent years, and the general problem is 
NP-hard \cite{katz2017reluplex}.  A large body of research focuses on the robustness issue of neural networks. In particular, given an input 
subject to (adversarial) perturbations, one intends to determine whether the output of the neural network (e.g., the classification result) is invariant to these perturbations. 
Essentially, this is to estimate the output range of a given neural network on a compact set.  
There are now a wide range of methods including constraint-solving based approaches \cite{katz2017reluplex}, 
optimization based approaches \cite{Dutta2018,WengZCSHDBD18,WeimingXiang2017}, 
abstract interpretation based approaches \cite{Pulina2010CAV,Chenlq19SAS}, etc.
Furthermore, recently work has been done for verification of control systems with neural network 
components \cite{DBLPSouradeep19HSCC,IvanovWAPL19,SunKS19,SESHIA19,NNV-CAV20}. The main technique is reachability analysis
of the closed-loop system, either by finite-state abstraction \cite{SunKS19}, or by interval-(or other abstract domain)-based 
reachable set approximation \cite{DBLPSouradeep19HSCC,IvanovWAPL19,NNV-CAV20}. Usually reachable set computation can only verify 
safety up to a finite time horizon, and the approximation error of reachable set may explode. Contrarily, we adopt the deductive 
approach based on barrier certificate, following and improving the line of work in \cite{Tuncali2018INVITEDRA}.

\subsection{Outline}
The rest of this paper is organized as follows: some preliminary knowledge is provided in Section~\ref{sec:pre} for self-containedness;
the main steps of our approach is presented in Section~\ref{sec:method} with a running example for demonstration; 
various improvements of the synthesized 
controllers are discussed in Section~\ref{sec:improve}; implementation and experiment details are given in Section~\ref{sec:implement}; the paper 
is concluded by Section~\ref{sec:conclude}. We note that a preliminary version is accepted by SETTA 2020 as a short paper under the same title. 


\section{Preliminaries}\label{sec:pre}

Throughout this paper,  $\mathbb R$ denotes the set of real numbers. For any natural number $n$, let $[n]=\{1, \cdots, n\}$.

\subsection{Constrained Continuous Dynamical System}
A continuous dynamical system is modeled by a system of
first-order ordinary differential equations (ODEs)
$\dot{\mathbf x}=\mathbf f(\mathbf x)$,
where 
\begin{itemize}
	\item $\mathbf x=(x_1,x_2, \ldots, x_n)^T\in\mathbb R^n$ is a column vector, $\dot{\mathbf x}$ denotes the derivative of $\mathbf x$ with respect to
	the time variable $t$, and 
	\item $\mathbf f(\mathbf x):\Omega \rightarrow \mathbb R^n$ is a vector field
	$\mathbf f(\mathbf x)=(f_1(\mathbf x),\cdots, $ $f_n(\mathbf x))^T$ defined on an
	open subset $\Omega \subseteq \mathbb R^{n}$.
\end{itemize} 
We assume that $\mathbf f$ satisfies
the \emph{local Lipschitz condition}, which ensures that, given $\mathbf x=\mathbf x_0$,
there exists a time $\mathcal T>0$ and a unique time trajectory $\mathbf x(t):
[0,\mathcal T)\rightarrow \mathbb R^{n}$ such that $\mathbf x(0)=\mathbf x_0$. In the sequel, the trajectory is denoted by $\mathbf x(t, \mathbf x_0)$.

A \emph{constrained continuous dynamical systems} (CCDS) is represented by $\Gamma=(\mathbf f, X_D, X_I, X_U)$,
where 
\begin{itemize}
	\item $\mathbf f: \Omega \rightarrow \mathbb R^n$ is the vector field, 
	\item $X_D\subseteq \Omega$ is an evolution constraint (or system domain),
	\item $X_I\subseteq X_D$, and 
	\item $X_U\subseteq X_D$. 
\end{itemize}
For CCDSs, the following problem is widely investigated in safety critical applications.

\begin{definition}[Safety Verification]\label{prob:safety-veri}
A CCDS $\Gamma=(\mathbf f, X_D, X_I, X_U)$ is safe
if $\forall \mathbf x_0 \in X_I$ and $\forall t \geq 0, \mathbf x(t, \mathbf x_0) \in X_D $ implies $\mathbf x(t, \mathbf x_0)\notin X_U$,
i.e., the system never reaches $X_U$ from $X_I$.
\end{definition}

\subsection{Controlled CCDS}
In this paper, we consider \emph{controlled CCDS} $\Gamma=(\mathbf f, X_D, X_I, X_U)$ with continuous dynamics defined by
\begin{equation}\label{eqn:control-system}
    \left\{
        \begin{array}{l}
            \dot {\mathbf x} =\mathbf  f(\mathbf x, \mathbf u)\\
            \mathbf u = \mathbf g(\mathbf x)
        \end{array}
        \right.,
\end{equation}
where $\mathbf x\in \mathbb R^n$, $\mathbf u\in U \subseteq \mathbb R^m$ are the feedback control inputs,
and $\mathbf f: \mathbb R^{n+m}\rightarrow \mathbb R^n$ and $\mathbf g:\mathbb R^n\rightarrow \mathbb R^m$
are locally Lipschitz continuous. The problem we considered in this paper is
defined as follows.
\begin{definition}[Safe Controller Synthesis]\label{prob:safe-synthesis}
    Given a controlled CCDS $\Gamma=(\mathbf f, X_D, X_I, X_U)$ with $\mathbf f$ defined by (\ref{eqn:control-system}),
    design a locally continuous feedback control law $\mathbf g$
    such that the closed-loop system $\Gamma$ with $\mathbf f = \mathbf f(\mathbf x, \mathbf g(\mathbf x))$ is safe,
    i.e. the system never reaches $X_U$ from $X_I$ under control $\mathbf u =\mathbf g(\mathbf x)$.
\end{definition}

\subsection{Barrier Certificate}  
Given a system $\Gamma$, a barrier certificate is a real-valued function $B(\mathbf x)$ over the states of the system satisfying 
the condition that $B(\mathbf x)\leq 0$ for any reachable state $\mathbf x$ and $B(\mathbf x) > 0$ for any
state in the unsafe set $X_U$. 
If such a function $B(\mathbf x)$ exists, one can easily deduce that 
the system can \emph{not} reach a state in the unsafe set from the initial set \cite{prajna2007safety,Ratschan18-tac}. 
In this paper, we will certify the safety of a synthesized 
controller by generating barrier certificates.

There are several different formulations of barrier certificates without explicit reference to the solutions of the ODEs 
\cite{prajna2007safety,kong2013exponential,dai2013barrier,Sogokon2018FM}. we will adopt what are called 
\emph{strict barrier certificate} \cite{sloth2012compositional} conditions.
\begin{theorem}[Strict barrier certificate] \label{thm:barrier}
	Given a system $\Gamma=(\mathbf f, X_D, X_I, X_U)$, if there exists a continuously differentiable function $B: X_D\rightarrow \mathbb R$ s.t.
	\begin{enumerate}
		\item $B(\mathbf x)\leq 0$ for $\forall \mathbf x \in X_I$
		\item $B(\mathbf x)>0$ for $\forall \mathbf x\in X_U$
		\item $\mathcal L_{\mathbf f}B(\mathbf x) <0 $ for all $\mathbf x \in X_D$ s.t. $B(\mathbf x)=0$,
	\end{enumerate}
then the system $\Gamma$ is safe, and such $B$ is a barrier certificate.
\end{theorem}
Note that in the above third condition, $\mathcal L_{\mathbf f} B$ is the \emph{Lie derivative} of $B$ w.r.t. $\mathbf f$, that is, the inner product 
of $\mathbf f$ and the gradient of $B$:
\begin{equation}\label{eqn:lie-def}
    \mathcal{L}_{\mathbf f} B(\mathbf x)=(\nabla B)\cdot
    \mathbf f(\mathbf x)=\sum_{i=1}^n\bigg(\frac{\partial B}{\partial x_i}(\mathbf x)\cdot
    f_i(\mathbf x)\bigg)\,.
\end{equation}

\subsection{Neural Networks}
\begin{figure}
    \centering
    \includegraphics[width=0.7\textwidth]{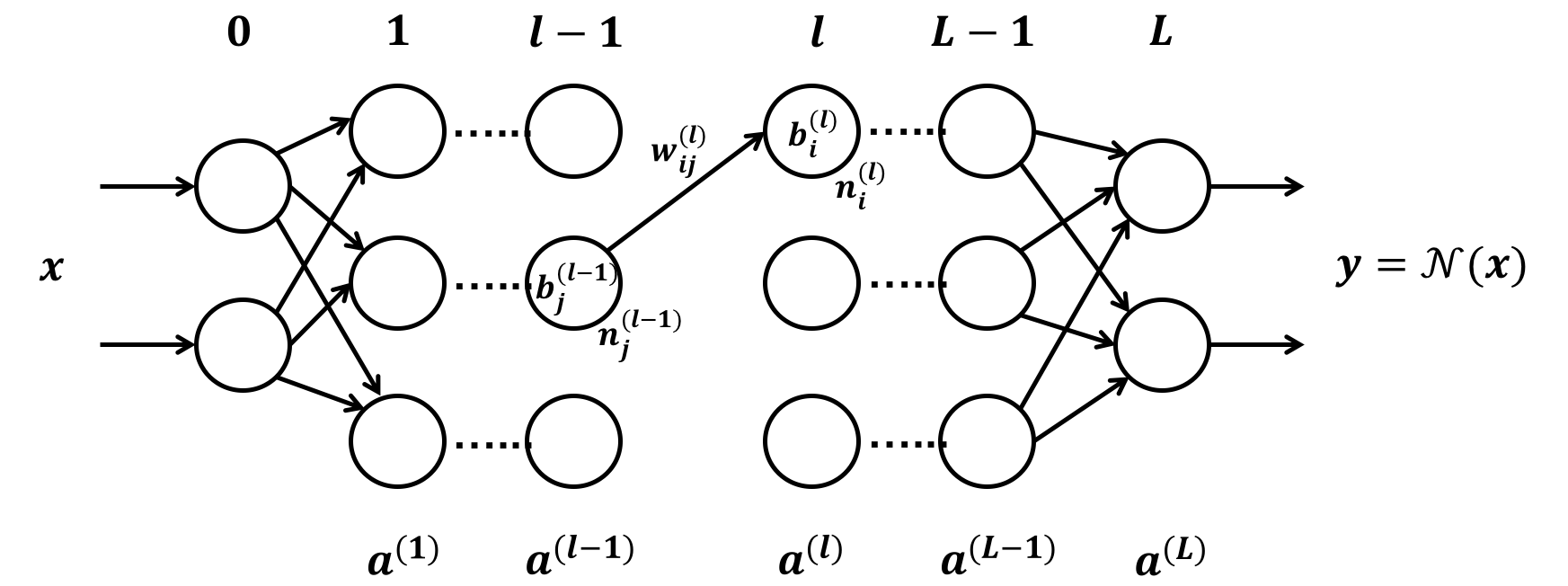}
    \caption{The structure of a multilayer feed-forward artificial neural network}\label{fig:NN}
\end{figure}

In this paper, both the synthesized control law $\mathbf g$ and the barrier certificate $B$ are represented by
(feed-forward artificial) neural networks (NNs).
We introduce some basic notions here. A typical NN consists of a number of
interconnected neurons which are organized in a layered structure. Each neuron is a single processing element that responds to the weighted inputs received from other neurons  (cf. Fig.~\ref{fig:NN}.)

In general, an NN represents a function $\mathcal N(\mathbf x)$ on the input $\mathbf x$ and can be represented as a composition of its layers. 
We normally reserve $0$ and $L$ for the indices of the input and the output layer respectively, and
all of the other layers in between are hidden layers. In this paper, 
we use superscripts to index layer-specific variables. In particular, 
the layer  $l$ comprises neurons $n_i^{(l)}$ for $i\in [d^{(l)}]$, where $d^{(l)}$ is the dimension of the layer $l$. 
Neuron $n_j^{(l-1)}$ of the layer $l-1$ is connected with neuron $n_i^{(l)}$ of layer $l$ by a directed edge with weight $w_{ij}^{(l)}\in\mathbb R$.
Each neuron $n_i^{(l)}$ of layer $l\in [L]$ is associated with a \emph{bias} $b^{(l)}_i\in \mathbb R$ and an activation function $a_i^{(l)}: \mathbb R\rightarrow \mathbb R$.
Usually the neurons in the same layer has identical activation functions, denoted by $a^{(l)}$.
Commonly used activation functions include ReLU  (rectified linear unit, i.e., $\max(0, x)$ for $x\in \mathbb R$), sigmoid, hyperbolic tangent, etc. 

Denote the input vector to the NN by $\mathbf x\in \mathbb R^{d^{(0)}}$. Let the output vector of the $l$-th layer be $\mathbf x^{(l)}$. 
Then $\mathbf x^{(0)}=\mathbf x$.
We introduce the vector variable $\mathbf z^{(l)}$ to denote the input vector to the $l$-th layer for $l\in[L]$.
Thus the \emph{forward propagation equations} of an NN can be
defined as
\begin{equation}\label{eqn:forward}
\left\{
\begin{array}{lllr}
\mathbf x^{(0)} &= & \mathbf x &\\
\mathbf z^{(l)} &= & \mathbf W^{(l)} \cdot \mathbf x^{(l-1)} + \mathbf b^{(l)} & \textrm{for } l\in[L]\\
\mathbf x^{(l)} &= & a^{(l)}(\mathbf z^{(l)}) & \textrm{for }  l\in[L]\\
\mathbf y &=  &\mathcal N(\mathbf x) = \mathbf x^{(L)}
\end{array} \right.,
\end{equation}
where $\mathbf W^{(l)}$ is a matrix of dimension $d^{(l)}\times d^{(l-1)}$,
$\mathbf b^{(l)}$ is a $d^{(l)}$-dimensional column vector, and
$a^{(l)}$ is taken as an element-wise function for a vector input.

Training of NN is usually through \emph{backward propagation}, 
during which the parameters $\mathbf W$'s and $\mathbf b$'s are learned through an 
optimization algorithm (e.g., stochastic gradient descent, SGD for short) applied on the training set \cite{dl-ian}.

\section{Methodology}\label{sec:method}
\begin{figure}
  \centering
	\includegraphics[width=0.8\textwidth]{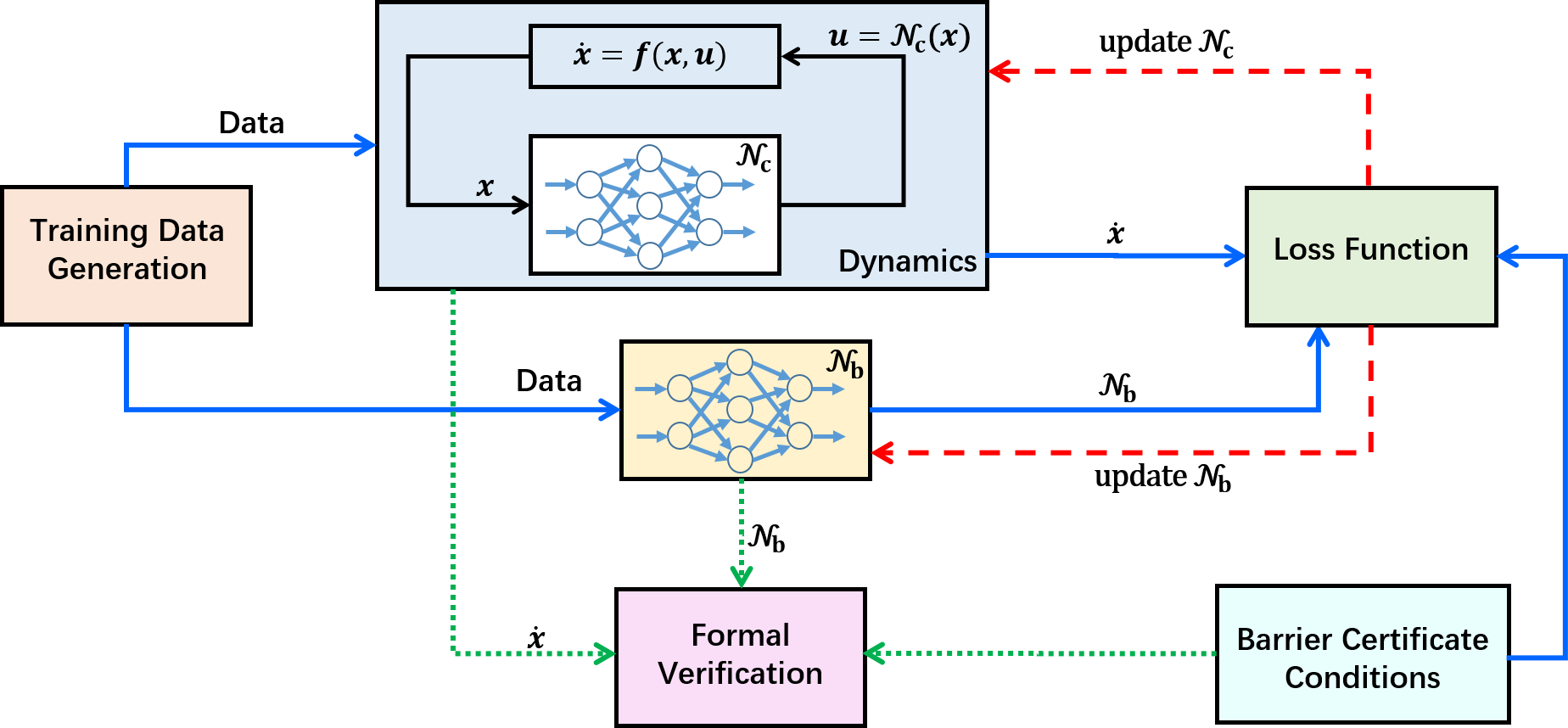}
	\caption{The framework of safe neural network controller synthesis} \label{fig:framework}
\end{figure}

The framework of our safe controller learning approach is demonstrated in 
Fig.~\ref{fig:framework}.
Given a controlled CCDS $\Gamma=(\mathbf f, X_D, X_I, X_U)$, the basic idea of the proposed approach is to represent the controller function $\mathbf g$ as well as the safety certificate function
$B$ by two NNs, i.e. $\mathcal N_{\rm c}$ and $\mathcal N_{\rm b}$ respectively. Then we formulate the barrier certificate conditions as per Theorem~\ref{thm:barrier} w.r.t.
$\mathcal N_{\rm b}$ and the closed-loop dynamics $\mathbf f(\mathbf x,\mathcal N_{\rm c}(\mathbf x))$ into a loss function, and then train the two NNs together
on a generated training data set until the loss is decreased to 0. The resulting two NNs are the controller and barrier certificate candidates. To overcome the 
limitations of data-driven approach, formal verification (SMT solver in this paper) is performed on the synthesized candidates to show that
the barrier certificate conditions are indeed satisfied. The blue (solid), red (dashed), and green (dotted) arrows in Fig.~\ref{fig:framework}
shows the information flow of forward propagation, backward propagation, and formal verification, respectively. Next, before giving more detailed steps 
of our approach, we first introduce a running example.

\begin{figure}
  \centering
  \begin{minipage}[b]{0.38\linewidth}
      \includegraphics[width=\textwidth]{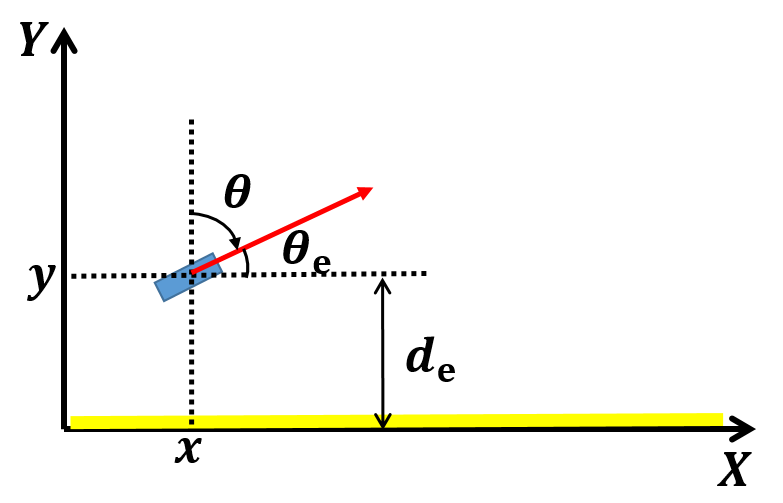}
      \caption{States of Dubins' car: $d_{\rm e}=y$, $\theta_{\rm e} =\frac{\pi}{2}-\theta$} \label{fig:dubins}
  \end{minipage}
  \quad
  \begin{minipage}[b]{0.58\linewidth}
      \includegraphics[width=\textwidth]{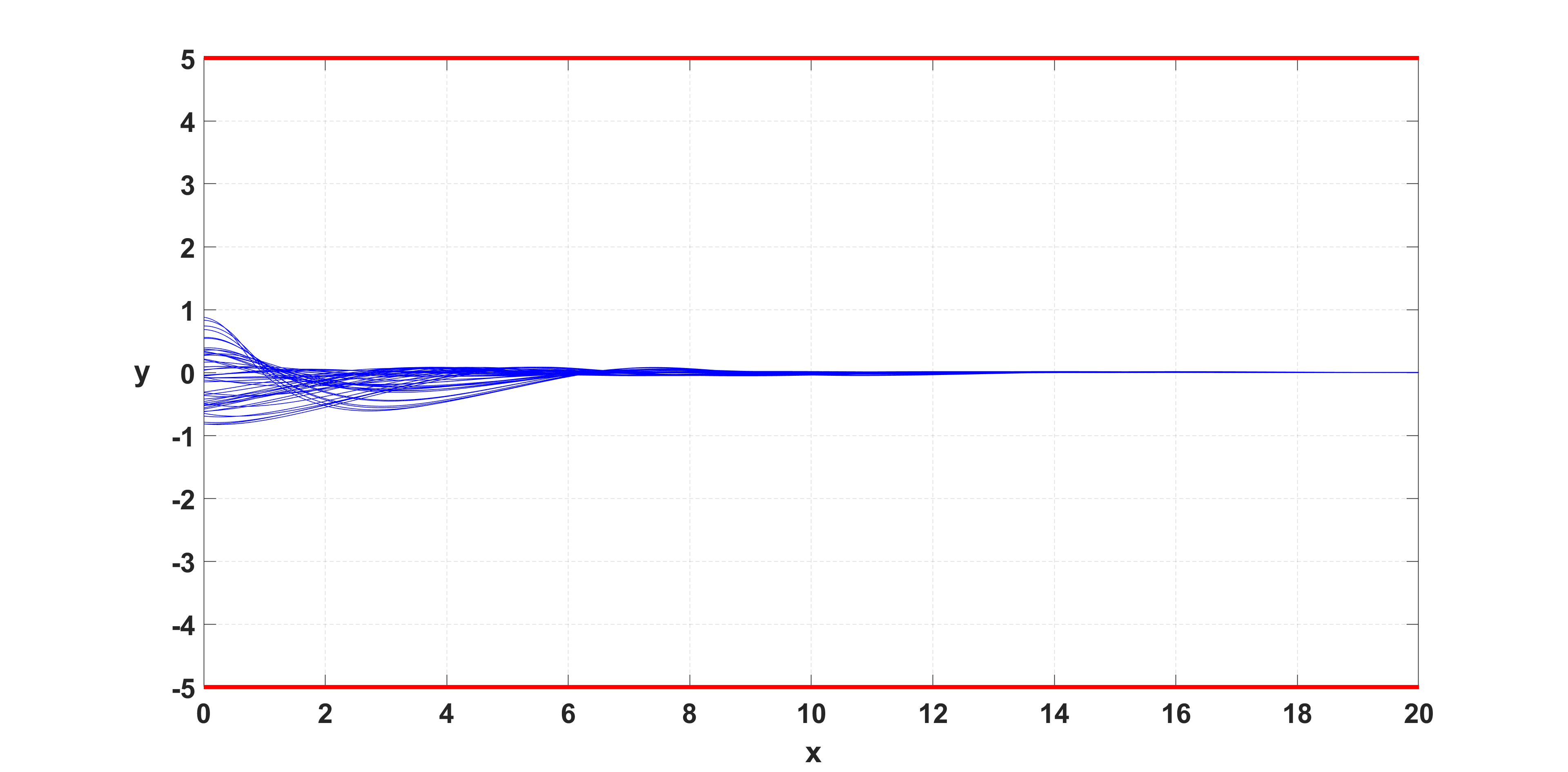}
      \caption{Simulated car trajectories with learned NN controller} \label{fig:dubins-sim}
  \end{minipage}
\end{figure}

\begin{example}[Dubins' Car \cite{Tuncali2018INVITEDRA,Deshmukh2019LearningDN}]\label{eg:dubins}
The control objective is to steer a car with constant velocity 1 to track a path, here the $X$-axis in the postive direction.
The states of the car are the $x,y$ position and the driving direction $\theta$, which can be transformed to the distance error $d_{\rm e}$ and 
angle error $\theta_{\rm e}$ between the current position and
the target path (see Fig.~\ref{fig:dubins}). The controlled CCDS $\Gamma = (\mathbf f, X_D, X_I, X_U)$ is:
\begin{equation}\nonumber
\mathbf f:
\left[
    \begin{array}{l}
    \dot{d_{\rm e}} \\ \dot{\theta_{\rm e}}
    \end{array}
\right] =
\left[
    \begin{array}{c}
     \sin(\theta_{\rm e}) \\-u
    \end{array}\right], \quad \textrm{where } u \textrm{ is the scalar control input}
\end{equation}
\begin{itemize}
\item $X_D$: $\{(d_{\rm e}, \theta_{\rm e})\in \mathbb R^2 | -6\leq d_{\rm e}\leq 6, -7 \pi/10\leq \theta_{\rm e} \leq 7 \pi/10\}$;
\item $X_I$: $\{(d_{\rm e}, \theta_{\rm e})\in \mathbb R^2 | -1\leq d_{\rm e}\leq 1, -\pi / 16\leq \theta_{\rm e}\leq \pi / 16\}$;
\item $X_U$: the complement of $\{(d_{\rm e}, \theta_{\rm e})\in \mathbb R^2 |  -5\leq d_{\rm e}\leq 5, -\pi / 2\leq \theta_{\rm e}\leq \pi / 2\}$ in $X_D$.
\end{itemize}
Figure~\ref{fig:dubins-sim} shows 50 simulated trajectories on the $x$-$y$ plane from random initial states in $X_I$ using our learned NN controller $u$.
The two red horizontal lines are the safety upper and lower bounds ($\pm 5$) for $y$ (the same bounds as $d_{\rm e}$).
In the rest of this paper, we will use Example~\ref{eg:dubins} to demonstrate our safe controller synthesis approach.
\end{example}

\subsection{The Structure of $\mathcal N_{\rm c}$ and $\mathcal N_{\rm b}$}\label{sec:nn-struc}

We first fix the structure of $\mathcal N_{\rm c}$ and $\mathcal N_{\rm b}$ as follows,
assuming that in the controlled CCDS $\Gamma$, $\mathbf x$ and $\mathbf u$ are of $n$ and $m$ dimension respectively, e.g. $n=2, m=1$ for Example~\ref{eg:dubins}.
\begin{itemize}
    \item {\bf Input layer} has $n$ neurons for both $\mathcal N_{\rm c}$ and $\mathcal N_{\rm b}$;
    \item {\bf Output layer} has $m$ neurons for $\mathcal N_{\rm c}$ and one single neuron for $\mathcal N_{\rm b}$;
    \item {\bf Hidden layer}: there is no restriction on the number of hidden layers or the number of neurons in each hidden layer;
    for Example~\ref{eg:dubins}, the structures are fixed as: $\mathcal N_{\rm c}$ has one hidden layer with 5 neurons, and $\mathcal N_{\rm b}$ has one hidden layer with 10 neurons;
    \item {\bf Activation function}: considering the inherent requirement of local Lipschitz continuity for $\mathcal N_{\rm c}$ and the inherent requirement of differentiability for $\mathcal N_{\rm b}$,
    and considering the simplicity of formal verification, 
    we adopt ReLU, i.e. $a(x)=\max(0,x)$, and {\emph{Bent-ReLU}} \cite{ZhaoZC020}, i.e., 
    \begin{equation}\label{eqn:bent-relu}
        a(x)=0.5\cdot x + \sqrt{0.25\cdot x^2 + 0.0001}
    \end{equation} as activation functions for hidden layers of $\mathcal N_{\rm c}$ and $\mathcal N_{\rm b}$ 
    respectively (the Lipschitz continuity of ReLU is by \cite{jordan2020exactly}); the activation function of the output layer is the identity map for 
both $\mathcal N_{\rm c}$ and $\mathcal N_{\rm b}$.
\end{itemize}

\subsection{Training Data Generation}\label{sec:data}
In our training algorithm, training data are generated by sampling points from the domain $X_D$, initial set $X_I$, and unsafe region $X_U$ of the considered system
$\Gamma$. No simulation of the continuous dynamics is needed. The  simplest sampling method is to grid the super-rectangles bounding
 $X_D$, $X_I$, $X_U$ with a fixed mesh size, and then filter out those points not satisfying the constraints of $X_D$, $X_I$, $X_U$. For example, 
we generate a mesh with $2^8\times 2^8$ points from $X_D$ for Example~\ref{eg:dubins}. The obtained three finite data sets are denoted by $S_D$, $S_I$, and $S_U$.

\subsection{Loss Function Encoding}\label{sec:loss}
Given $S_I$, $S_U$, and $S_D$, the loss function for training $\mathcal N_{\rm c}$ and $\mathcal N_{\rm b}$ can be expressed as
\begin{equation}\label{eqn:cost-full-data}
    L(S_D, S_I, S_U) = c_1 \cdot \sum_{\mathbf x\in S_I} L_1(\mathbf x) + c_2 \cdot \sum_{\mathbf x\in S_U} L_2(\mathbf x) + c_3\cdot \sum_{\mathbf x\in S_D} L_3(\mathbf x)  
\end{equation}
with 
\begin{equation}\label{eqn:cost-123}
    \begin{array}{ll}
    L_1(\mathbf x) = \text{ReLU}(\mathcal N_{\rm b}(\mathbf x) + \varepsilon_1) &\text{for } \mathbf x \in S_I\,, \\
    L_2(\mathbf x) = \text{ReLU}(-\mathcal N_{\rm b}(\mathbf x) + \varepsilon_2)&\text{for } \mathbf x \in S_U\,, \\
    L_3(\mathbf x) = 
           \text{ReLU}\big( \mathcal{L}_{\mathbf f}\mathcal N_{\rm b}(\mathbf x)+ \varepsilon_3\big) & 
           \text{for } \mathbf x\in \{\mathbf x\in S_D: |\mathcal N_{\rm b}(\mathbf x)|\leq \varepsilon_4\} \\
    \end{array}
\end{equation}
denoting the sub-loss functions encoding the three conditions of Theorem~\ref{thm:barrier}, and $c_1,c_2,c_3$ the three positive constant weight coefficients 
for the sub-losses $L_1,L_2,L_3$ respectively. 
The basic idea is to impose a positive (resp., zero) penalty to those sampled points
that violate (resp., satisfy) barrier certificate conditions. The $\varepsilon_1,\varepsilon_2, \varepsilon_3$ in (\ref{eqn:cost-123}) are three small non-negative
tolerances, the role of which is to get the non-sampled points around the sampled data to have zero loss as well. The $\varepsilon_4$ in (\ref{eqn:cost-123}) is a small positive constant characterizing 
a narrow belt region around the zero-level set of $\mathcal N_{\rm b}$, since we cannot sample data on the level set exactly. Note that in the 
above expression $L_3$, $\mathbf f$ is $\mathbf f(\mathbf x, \mathcal N_{\rm c}(\mathbf x))$.

\subsection{The Training Process}\label{sec:training}
We adopt a modified SGD optimization technique for training the two NNs $\mathcal N_{\rm c}$ and $\mathcal N_{\rm b}$. That is, we partition the training data sets $S_D,S_I,S_U$ into mini-batches and shuffle the list
of batches to gain some randomness effect, rather than shuffling the whole training data set. For each mini-batch of data, the loss is calculated according to 
(\ref{eqn:cost-full-data}) and the the weights and biases of the two NNs are updated by a gradient descent step through backward propagation. To start the training, we must first specify the $\varepsilon_1$ to $\varepsilon_4$
in the loss function, as well as hyper-parameters such as number of restarts $n_{\rm restart}$, number of epoches $n_{\rm epoch}$, number of mini-batches $n_{\rm batch}$, and learning rate $l_{r}$, etc. 
For Example~\ref{eg:dubins}, we set $n_{\rm restart}=5$, $n_{\rm epoch}=100$, $n_{\rm batch}=4096$ and $l_{r}=0.1$. The choices of 
$\varepsilon_1$ to $\varepsilon_4$ will be presented in the following subsection. The training process terminates when the loss is decreased to 0 on all
mini-batches or the number of restarts exceeds $n_{\rm restart}$.

\subsection{Formal Verification}\label{sec:verification}
The rigorousness of the NNs resulted from 0 training loss is not guaranteed since our approach is data-driven, that is, the three conditions in Theorem~\ref{thm:barrier} are
not necessarily satisfied by $\mathcal N_{\rm c}$ and $\mathcal N_{\rm b}$. Therefore we resort to formal verification to 
guarantee the correctness our synthesized controllers. To preform the verification, we replace $\mathbf f$ and $B$ in the conditions of Theorem~\ref{thm:barrier}
by $\mathbf f(\mathbf x, \mathcal N_{\rm c}(\mathbf x))$ and $\mathcal N_{\rm b}$, and try to show that the negation of the conjunction of the three conditions, i.e.
\begin{equation}\label{eqn:veri-condition}
    \begin{array}{ll}
    &\exists\mathbf x.\, \mathbf x\in X_I \wedge \mathcal N_{\rm b}(\mathbf x) > 0\\
    \vee&\exists\mathbf x.\, \mathbf x\in X_U \wedge \mathcal N_{\rm b}(\mathbf x) \leq 0\\
    \vee&\exists\mathbf x.\, \mathbf x\in X_D \wedge \mathcal N_{\rm b}(\mathbf x) = 0 \wedge \mathcal L_{\mathbf f(\mathbf x, \mathcal N_{\rm c}(\mathbf x))} \mathcal N_{\rm b}(\mathbf x) \geq 0
    \end{array}
\end{equation}
is {\small UNSATISFIABLE}. Due to the high degree of nonlinearity in $\mathbf f$ and $\mathcal N_{\rm b}$ of (\ref{eqn:veri-condition}), its satisfiability is resolved by
the interval-propagation based nonlinear SMT solver {\sf iSAT3}.\footnote{\url{https://projects.informatik.uni-freiburg.de/projects/isat3/}}
To speed up the verification process, 
we compute piece-wise linear approximations (with interval error bounds) of Bent-ReLU function and its derivative, and replace their occurrences in $\mathcal N_{\rm b}$ and $\mathcal L_{\mathbf f}\mathcal N_{\rm b}$ by 
the linear approximations.
As a result, there are three issues that may affect the efficiency and effectiveness of formal verification:
\begin{itemize}
    \item The tolerances chosen for loss function encoding in (\ref{eqn:cost-full-data}) and (\ref{eqn:cost-123});
    \item The piece-wise linear approximation error of Bent-ReLU function and its derivative;
    \item The interval splitting width for \sf{iSAT3}.
\end{itemize}
For the third issue, we usually set the minimal splitting width option {\sf -}{\sf -msw} to 0.001 for {\sf iSAT3}. The first and second issues are addressed in the following two paragraphs.

\begin{figure}
  \centering
  \includegraphics[width=0.6\textwidth]{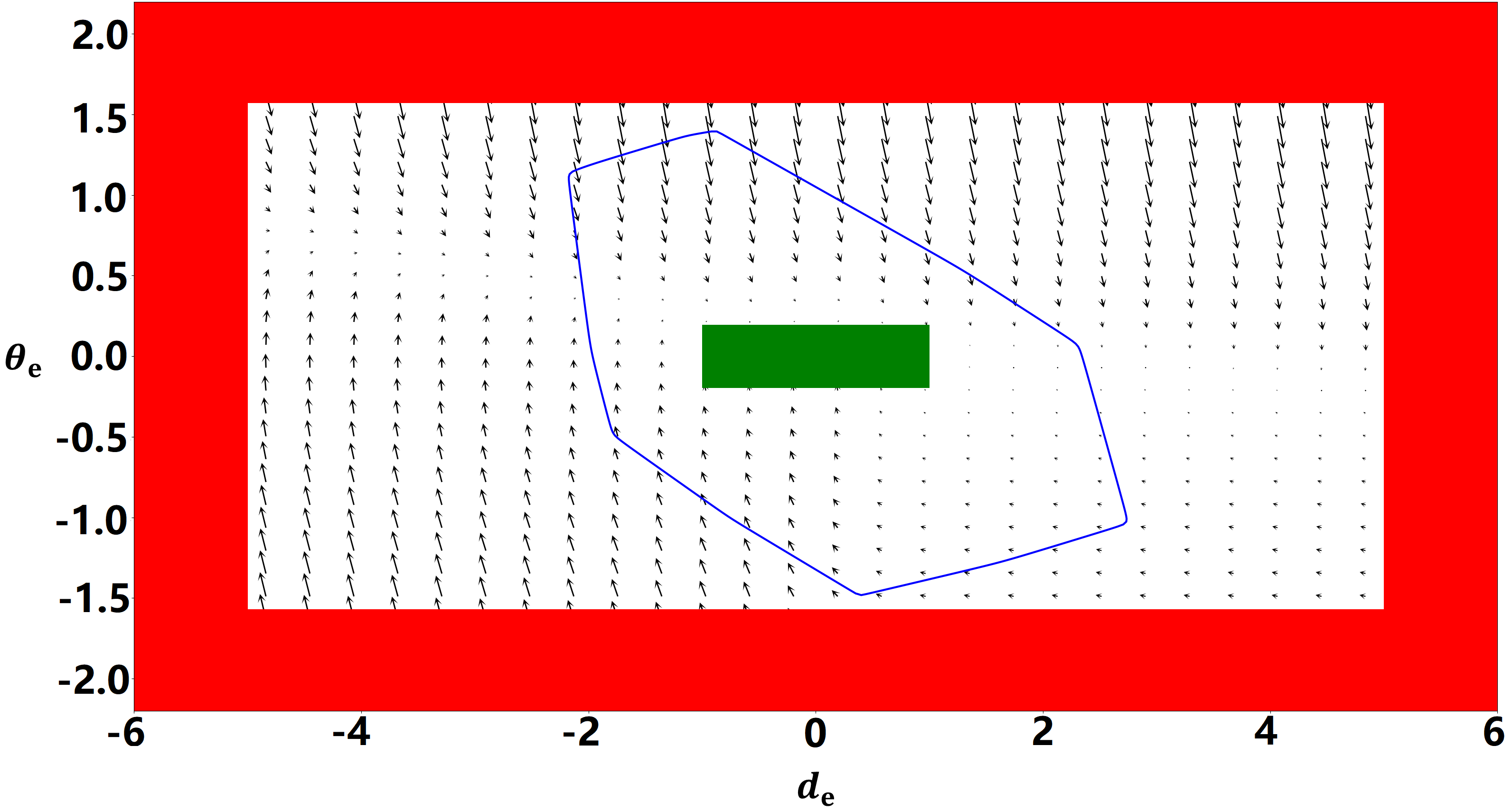}
  \caption{Learned and verified NN controller and barrier certificate for Example~\ref{eg:dubins}: 
  the inner (green) and outer (red) shaded areas are the initial and unsafe regions, black arrows in the white area are the closed-loop vector fields $\mathbf f(\mathbf x, \mathcal N_{\rm c}(\mathbf x))$, 
  and the blue curve surrounding the inner shaded box is the zero-level set of $\mathcal N_{\rm b}$
  }\label{fig:barriers}
\end{figure}

\paragraph{Pre-training and Fine-tuning.} The success of synthesis and formal verification heavily relies on the choices of the four constants $\varepsilon_1$ to $\varepsilon_4$ 
in (\ref{eqn:cost-full-data}) and (\ref{eqn:cost-123}). Generally, small tolerances are preferred for faster training, while larger tolerances are preferred 
for formal verification to compensate for the errors caused by activation function linearization and interval arithmetic computation.
In practice, we adopt a pre-training and fine-tuning combination strategy. That is, we start with small positive $\varepsilon_4$ and zero
$\varepsilon_1$ to $\varepsilon_3$ to perform the initial training. If the pre-trained NNs failed formal verification, they are iteratively refined by gradually increasing  
the tolerances.
For Example~\ref{eg:dubins}, the first controller and barrier
certificate are synthesized with $\varepsilon_4=0.01$ and $\varepsilon_1=\varepsilon_2=\varepsilon_3=0$, for which formal verification fails, while the fine-tuned controller and barrier certificate are successfully verified
when $\varepsilon_3$ was increased to $0.01$ (see Fig.~\ref{fig:barriers}).

\begin{figure}
  \centering
  \subfigure[Negative $\mathcal L_{\mathbf f}B$ with large $|\mathcal L_{\mathbf f}B|$]{
  \includegraphics[width=0.45\textwidth]{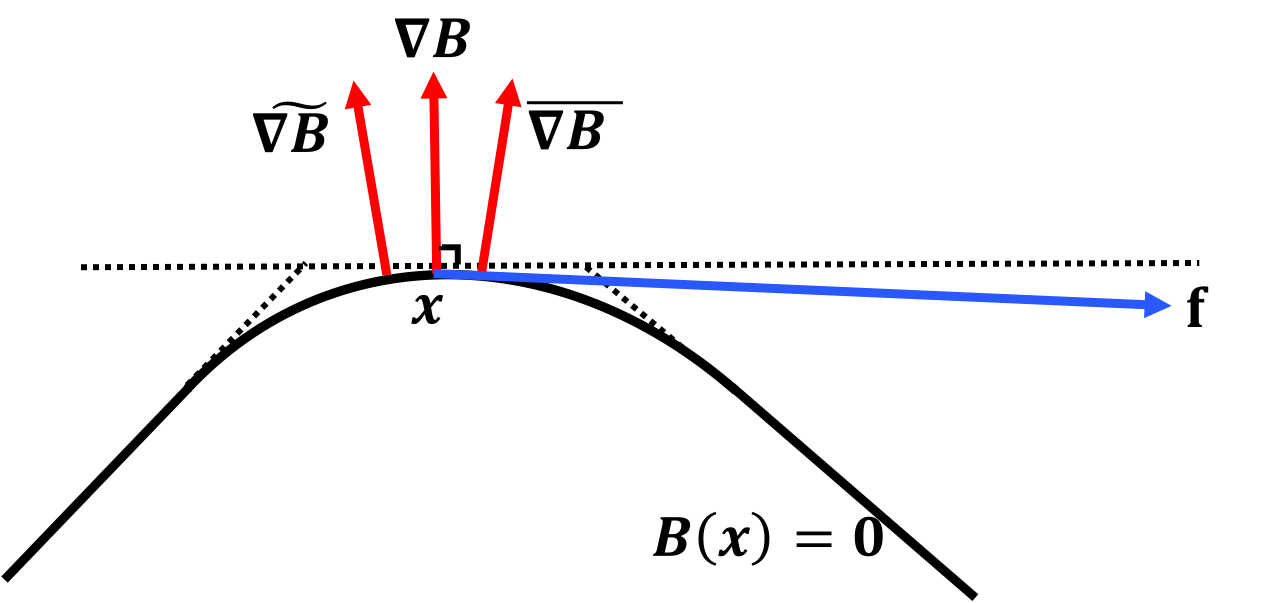}\label{fig:deri-error-a}}
  \qquad\qquad\,\,\,
  \subfigure[Negative $\frac{\mathcal L_{\mathbf f}B}{\|\nabla B\|\cdot\|\mathbf f\|}$ with large $\frac{|\mathcal L_{\mathbf f}B|}{\|\nabla B\|\cdot\|\mathbf f\|}$]{
  \includegraphics[width=0.36\textwidth]{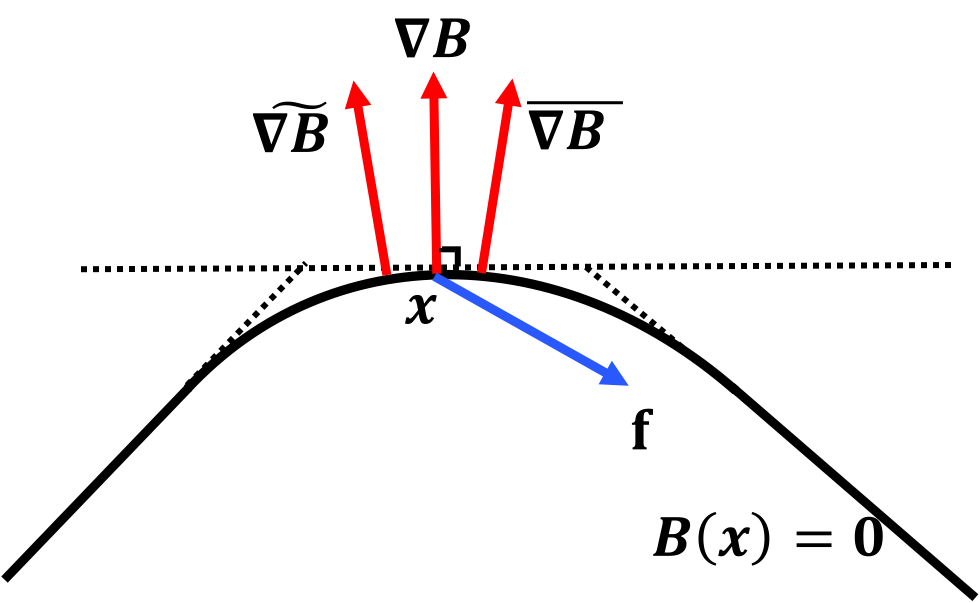}\label{fig:deri-error-b}}
  \caption{The sign of normalized Lie derivative is robust to Bent-ReLU linearization errors}\label{fig:deri-error}
\end{figure}

\paragraph{Adding Normalized Lie Derivative in Loss Encoding.}
Larger tolerances in the loss function (\ref{eqn:cost-full-data}) and (\ref{eqn:cost-123}) are not always useful for formal verification. To see this, consider checking unsatisfiability of the third condition of (\ref{eqn:veri-condition}).
Noting that $\mathcal L_{\mathbf f}B=\nabla B \cdot \mathbf f=\|\nabla B\|\|\mathbf f\|\cos{\theta_{\nabla B,\mathbf f}}$,
where $\|\cdot\|$ denotes the Euclidean norm and $\theta_{\nabla B, \mathbf f}$ denotes the angle between $\nabla B$ and $\mathbf f$,
Fig.~\ref{fig:deri-error-a} illustrates a situation that a point $\mathbf x$ on the zero-level set of a barrier candidate $B$ has negative Lie derivative, since
$\theta_{\nabla B, \mathbf f}$ is slightly larger than $\frac{\pi}{2}$ at $\mathbf x$. Moreover, it can be concluded that $\mathcal L_{\mathbf f}B(\mathbf x)<-\varepsilon_3$ 
for very large $\varepsilon_3$ since $\|f\|$ is large. However, formal verification of the negative Lie derivative condition would be very hard
at $\mathbf x$, where $\nabla B$ has a large approximation error due to linearization. For instance, if the approximated $\nabla B(\mathbf x)$ ranges from $\widetilde{{\nabla B}}$ to $\overline{\nabla B}$,
then formal verification becomes impossible since $\theta_{\overline{\nabla B},\mathbf f}<\frac{\pi}{2}$ which makes the Lie derivative positive. The reason for such a phenomenon 
is that negative $\mathcal L_{\mathbf f} B$ does not necessarily force the span angle of $\nabla B$ and $\mathbf f$ to be large, so the sign of $\mathcal L_{\mathbf f}B$ is not robust to
approximation noises of $\nabla B$. The problem can be resolved by introducing additional sub-loss function specifying normalized Lie derivative into the loss function (\ref{eqn:cost-full-data}) as follows:
\begin{equation}\label{eqn:cost-norm-lie}
    \begin{array}{ll}
    L_4(\mathbf x) = 
           \text{ReLU}\big(\frac{\mathcal{L}_{\mathbf f}\mathcal N_{\rm b}(\mathbf x)}{\|\nabla \mathcal N_{\rm b}\|\cdot\|\mathbf f\|}+ \varepsilon_5\big), & 
           \text{for } \mathbf x\in \{\mathbf x\in S_D: |\mathcal N_{\rm b}(\mathbf x)|\leq \varepsilon_4\} \\
    \end{array}
\end{equation}
where $\varepsilon_4$ are defined in (\ref{eqn:cost-123}) and $\varepsilon_5$ is a non-negative constant. By (\ref{eqn:cost-norm-lie}), if a barrier certificate is synthesized with
zero $L_4$ value and enough large $\varepsilon_5$, then the angle between $\nabla \mathcal N_{\rm b}$ and $\mathbf f$ would be large enough to tolerant large approximation errors of gradient (cf. Fig.~\ref{fig:deri-error-b}).

\begin{figure}[h!]
  \centering
  \includegraphics[width=0.8\textwidth]{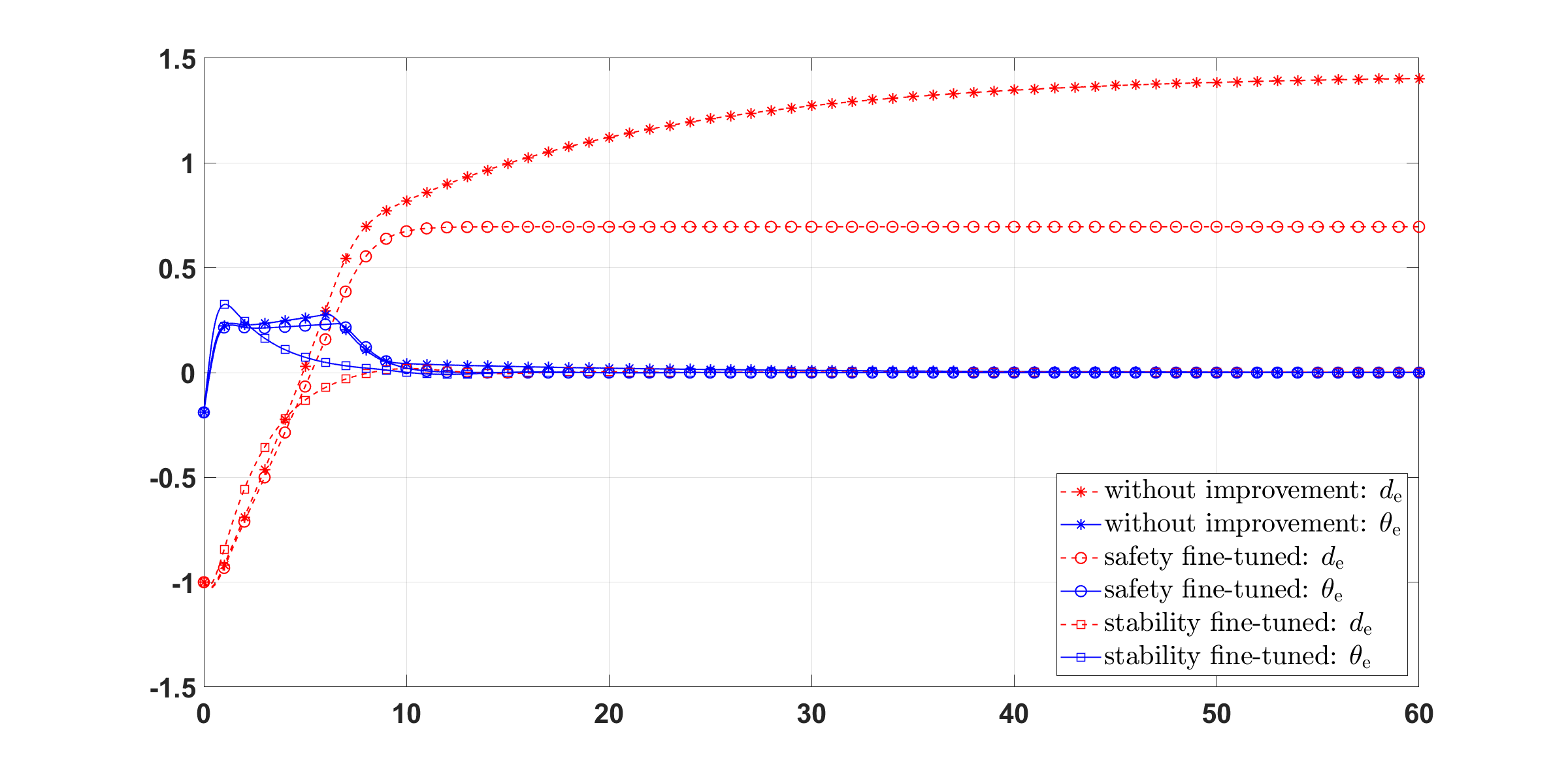}\
  \caption{Simulations of Dubins' car from $(-1,-0.19)$ with different NN controllers for comparison of stability performance}\label{fig:stability-compa}
\end{figure}
\section{Improvement of the Learned Controllers}\label{sec:improve}
The controller synthesized and verified in the last section is guaranteed to be safe. However, it may perform poorly regarding properties such as 
stability. As an illustration, we simulate the Dubins' car system from initial state $d_{\rm e}=-1,\theta_{\rm e}=-0.19$ using the NN controller corresponding to Fig.~\ref{fig:barriers}.
The changes of $d_{\rm e}$ and $\theta_{\rm e}$ within 60 time units are shown in Fig.~\ref{fig:stability-compa} by $\ast$-marked dashed ($d_{\rm e}$) or solid ($\theta_{\rm e}$) lines.
It is obvious that the car has a large distance error although it is still within safety bounds ($\pm 5$). 
We therefore propose a series of ways to improve the performance of synthesized controllers 
in this section.

\subsection{Larger Safety Margin}
\begin{figure}
  \centering
  \includegraphics[width=0.6\textwidth]{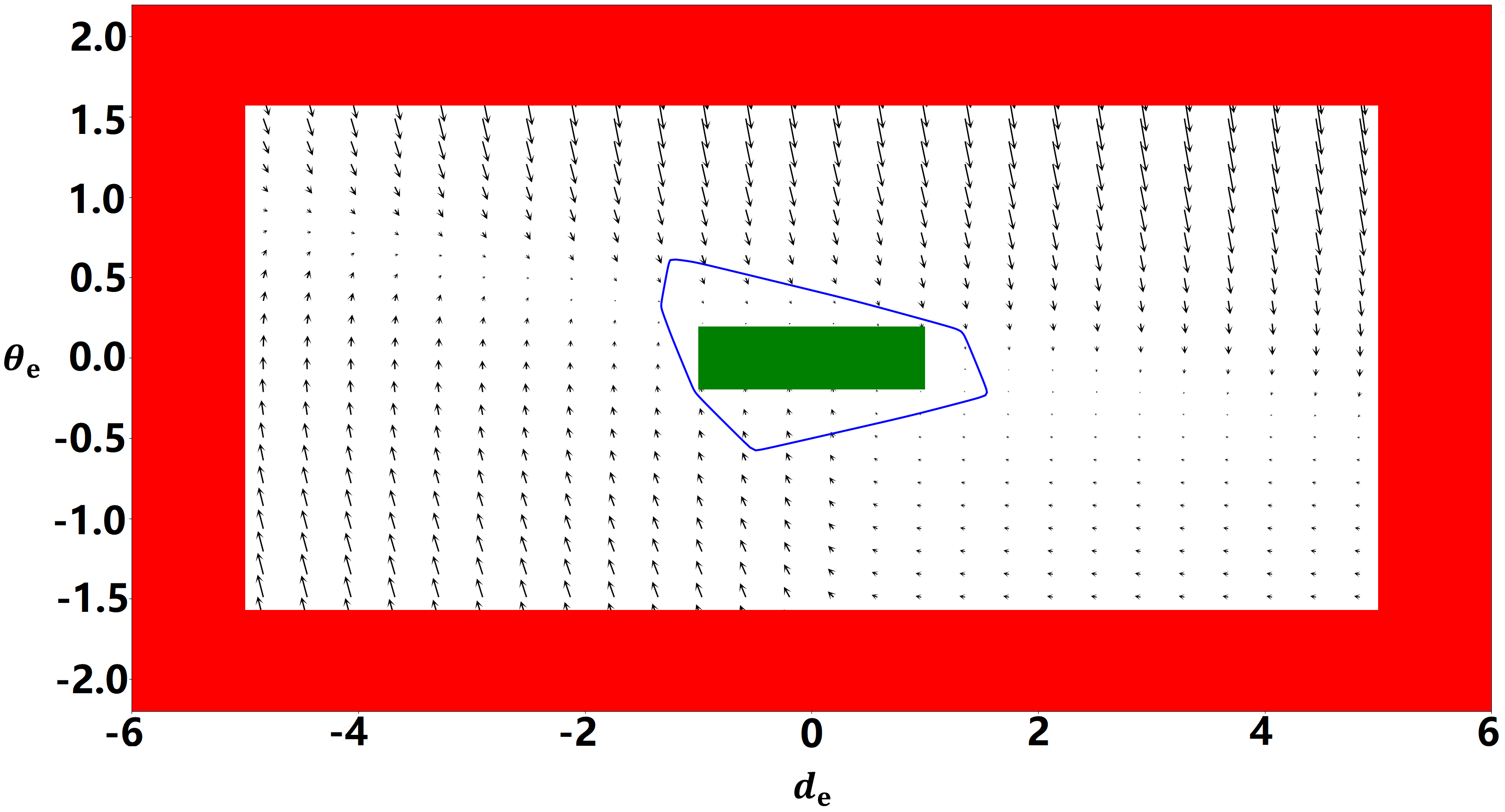}
  \caption{NN controller learned and verified for Example~\ref{eg:dubins} with larger safety margin: $\varepsilon_1 = 0.02$, $\varepsilon_2 = 0.8$, $\varepsilon_3 = 0.01$, $\varepsilon_4 = 0.05$; 
  the inner (green) and outer (red) shaded areas are the initial and unsafe regions, black arrows in the white area are the closed-loop vector fields $\mathbf f(\mathbf x, \mathcal N_{\rm c}(\mathbf x))$, 
  and the blue curve surrounding the inner shaded box is the zero-level set of $\mathcal N_{\rm b}$}
  \label{fig:large-s}
\end{figure}

The first improvement is to gradually increase the safety margin specified by the $\varepsilon_2$ constant in the loss function 
(\ref{eqn:cost-full-data}) and (\ref{eqn:cost-123}) by iterative fine-tuning. For example,
when $\varepsilon_2$ is increased to 0.8, a NN controller $\mathcal N_{\rm c}$ and the corresponding barrier $\mathcal N_{\rm b}$ are synthesized and 
shown in Fig.~\ref{fig:large-s}. The simulation performance of $\mathcal N_{\rm c}$ is shown in
Fig.~\ref{fig:stability-compa} by $\circ$-marked dashed ($d_{\rm e}$) or solid ($\theta_{\rm e}$) lines. 
It is obvious that distance error is reduced compared to the controller of Fig.~\ref{fig:barriers}.

\subsection{Asymptotic Stability}
\begin{figure}
  \centering
  \includegraphics[width=0.8\textwidth]{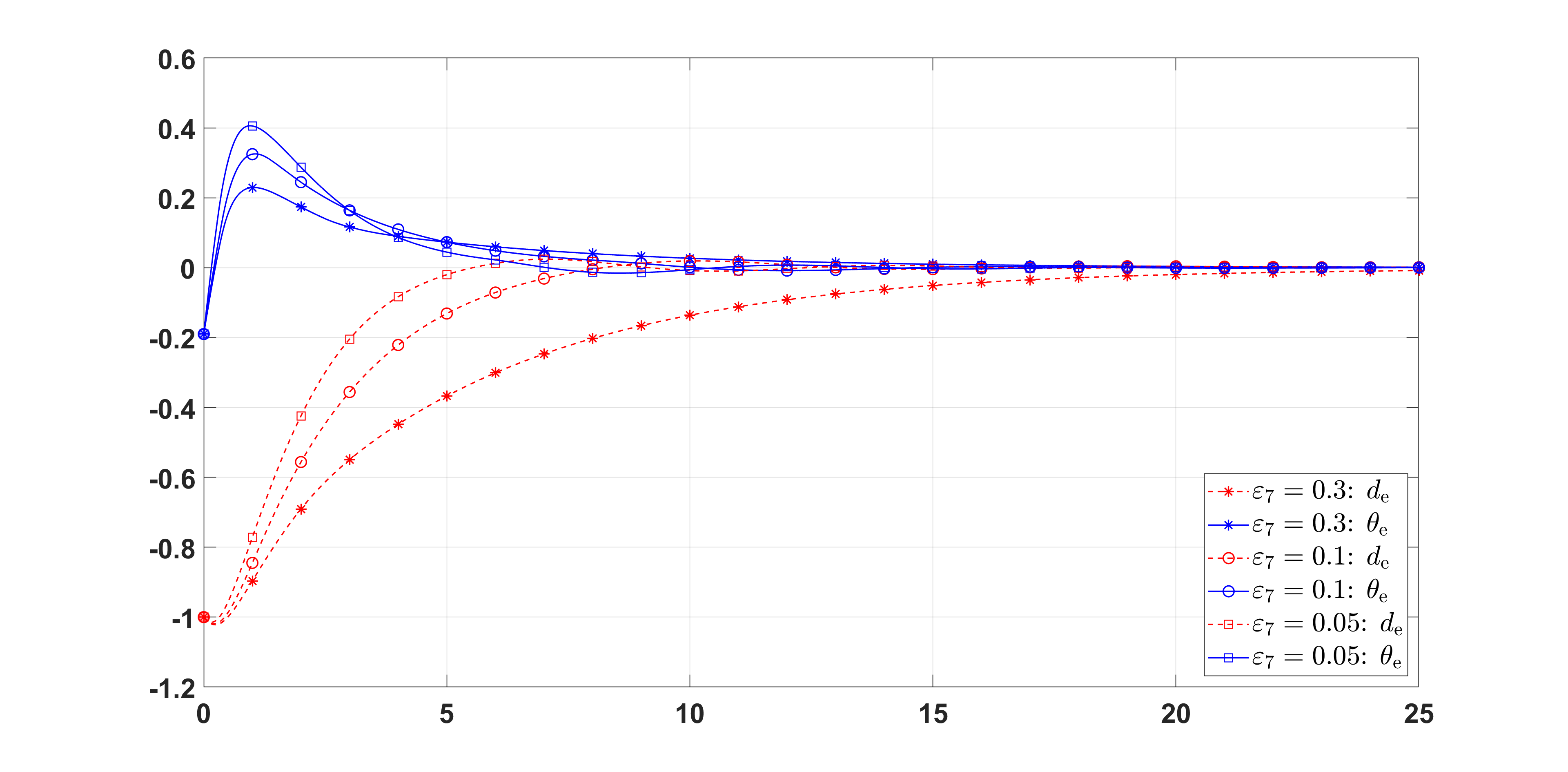}
  \caption{Comparison of NN controllers learned using $L_5$ and $L_6$ losses with $\varepsilon_6=0.05$, $\varepsilon_8=0.001$ for Example~\ref{eg:dubins}: 
  all simulations are from initial state $(-1, -0.19)$; dashed and solid lines represent $d_{\rm e}$ and $\theta_{\rm e}$ traces respectively; simulations corresponding to controllers learned with 
  $\varepsilon_7=0.3, 0.1, 0.05$ are marked by $*$, $\circ$, and $\scriptstyle{\square}$ respectively}
  \label{fig:para-effect}
\end{figure}

Figure~\ref{fig:stability-compa} shows that using the NN controller with larger safety margin, the distance error of the Dubins' car stabilizes at a value larger than 0.5, which is 
not desirable. To further reduce the distance error in the long time, we introduce additional loss terms into the loss function to express asymptotic-stability-like properties. 
Suppose that $\mathbf x_{o}$
is an expected equilibrium point of the system, that is, $\mathbf f(\mathbf x_o, \mathcal N_{\rm c}(\mathbf x_o))=\mathbf 0$. For example, the system in Example~\ref{eg:dubins} is expected to stabilize with 0 distance and angle errors
and so $\mathbf x_{o}$ is $(0,0)$.
Then we define the sub-loss functions for asymptotic stability as:
\begin{equation}\label{eqn:loss-stablity}
    \begin{array}{llll}
    L_5(\mathbf x) &=& \text{ReLU}\big(-\|\mathbf f(\mathbf x, \mathcal N_{\rm c}(\mathbf x))\|+ \varepsilon_6\big)& \text{for } \mathbf x \in \{\mathbf x\in S_D: \|\mathbf x-\mathbf x_{o}\|> \varepsilon_7\}\,,\\
    L_6(\mathbf x) &=& \text{ReLU}\big(\|\mathbf f(\mathbf x, \mathcal N_{\rm c}(\mathbf x))\| - \varepsilon_8\big) & \text{for } \mathbf x = \mathbf x_{o}
    \end{array}
\end{equation}
where $\varepsilon_6,\varepsilon_7,\varepsilon_8$ are three small non-negative constants. The basic idea of $L_5,L_6$ is to impose such constraints that the closed-loop vector 
field $\mathbf f(\mathbf x, \mathcal N_{\rm c}(\mathbf x))$ has negligible norm at the asymptotically stable point $\mathbf x_{o}$, 
and strictly positive norm outside a neighborhood of $\mathbf x_{o}$
with radius $\varepsilon_7$.
By choosing $\varepsilon_7 = 0.1$, $\varepsilon_6 = 0.05$, $\varepsilon_8 = 0.001$ we obtain a fine-tuned $\mathcal N_{\rm c}$ 
whose simulation performance is shown in
Fig.~\ref{fig:stability-compa} by $\scriptstyle{\square}$-marked dashed ($d_{\rm e}$) or solid ($\theta_{\rm e}$) lines, which demonstrate good asymptotic stability property. 
We also fix $\varepsilon_8=0.001, \varepsilon_6=0.05$ and compare the performances of $\mathcal N_{\rm c}$ obtained from different $\varepsilon_7$ values. The simulation 
results are shown in Fig.~\ref{fig:para-effect}. It can be roughly concluded that decreasing $\varepsilon_7$ will have an effect of increasing the \emph{overshoot} and decreasing the \emph{settling time}
of the simulated traces. An intuitive explanation of such effects is that by $L_5$, shrinking $\varepsilon_7$ increases $\|\mathbf f\|$ near $\mathbf x_{o}$, and thus trajectories approaches
$\mathbf x_{o}$ quickly but may overshoot. 

\begin{figure}
    \centering
    \includegraphics[width=0.8\textwidth]{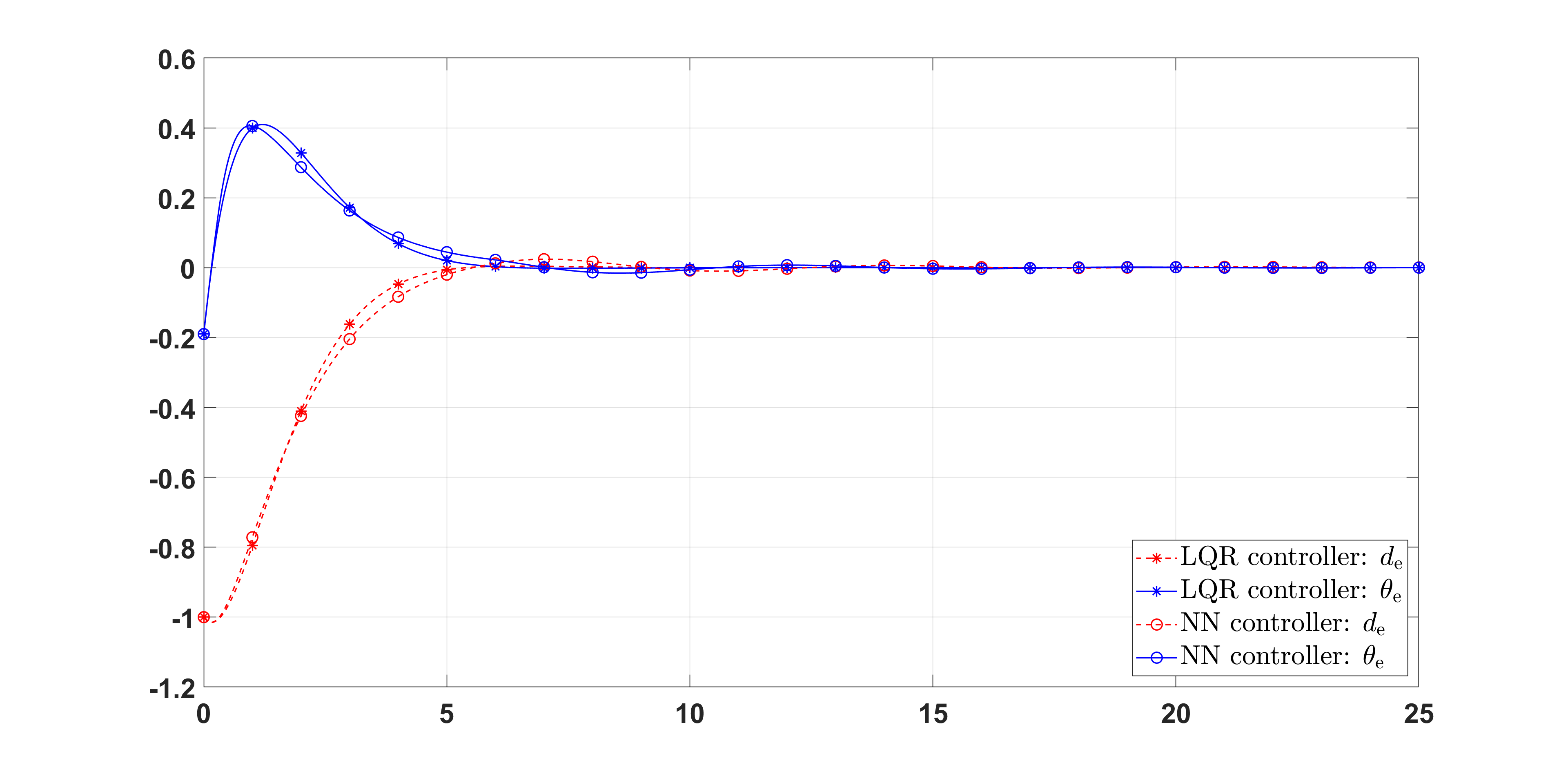}
    \caption{Simulation of NN and LQR controllers with initial state $(-1,-0.19)$ for Example~\ref{eg:dubins}: the NN controller is synthesized with 
    $\varepsilon_6=\varepsilon_7=0.05, \varepsilon_8=0.001$, and the LQR controller is synthesized with  
    $Q$ the 2-dimensional identity matrix and $R=1$; dashed and solid lines represent $d_{\rm e}$ and $\theta_{\rm e}$ traces respectively, and traces simulated with LQR and NN controllers are 
    marked by $*$ and $\circ$ respectively}\label{fig:lqr}
\end{figure}

\paragraph{Comparison with LQR Controllers.}
To further evaluate the performance of synthesized NN controllers, we linearize the Dubins' car system near $\mathbf x_{o}=(0,0)$ and then compute
the classic LQR (linear quadratic regulator \cite{linear-system-2}) controllers for the linearized system. 
Preliminary experiment shows that for fixed $Q$ and $R$ matrices in the LQR controller computation, by tuning the values of $\varepsilon_6$ and $\varepsilon_7$, 
we can obtain NN controllers with comparable performances to LQR controllers (cf. Fig.~\ref{fig:lqr}). 
\begin{remark}
    NNs controllers are in principle much more expressive than linear controllers such as LQR, and so it is interesting to 
    investigate better ways of loss function encoding and controller tuning to gain superior NN controllers to linear controllers (e.g. LQR) in 
    future.
\end{remark}

\subsection{Bounded Control Inputs}
\begin{figure}
    \centering
    \subfigure[Unbounded NN controller]{
    \includegraphics[width=0.45\textwidth]{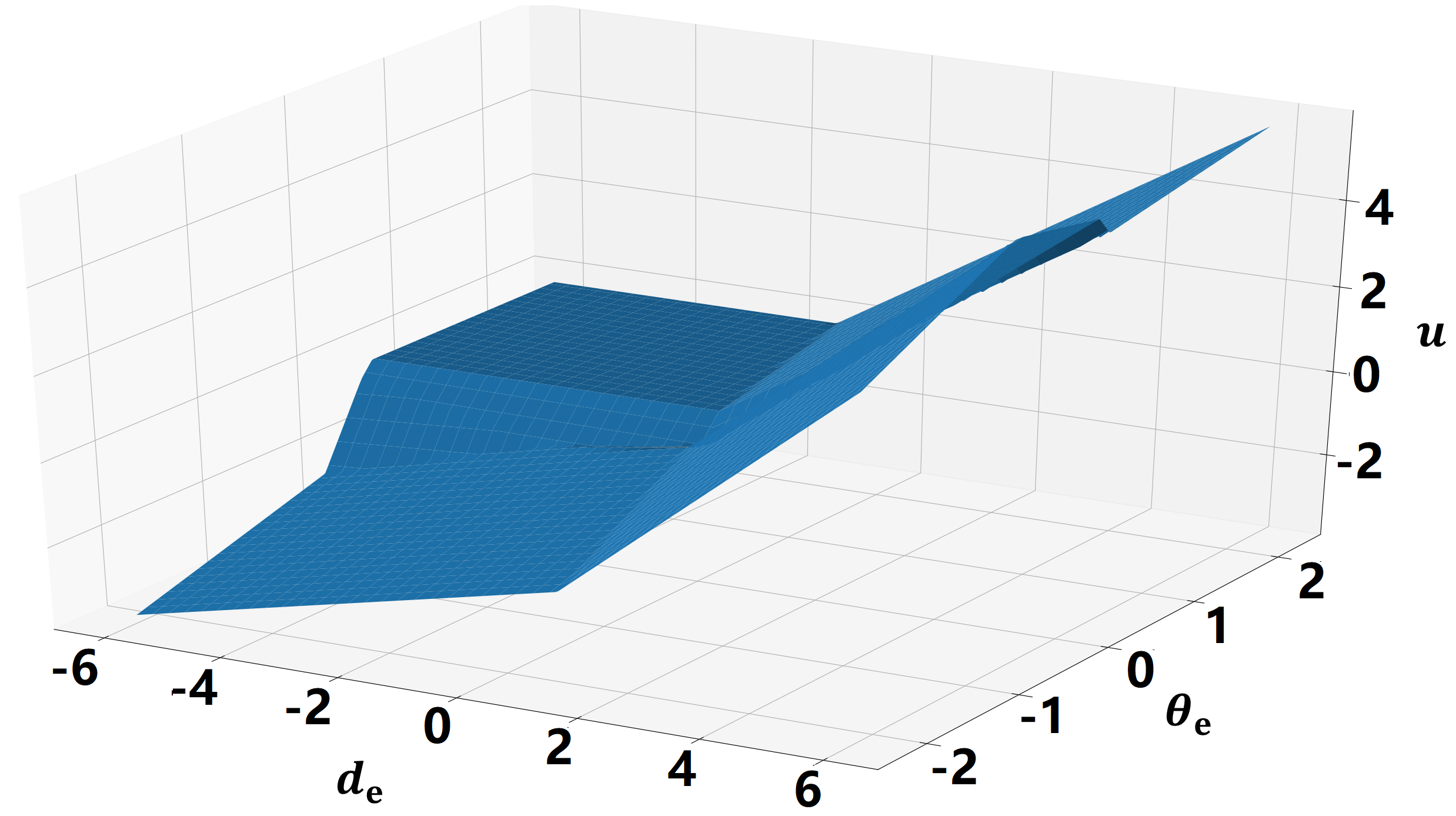}
    \label{fig:unbounded-sub}}
    \quad\,\,\,
    \subfigure[Bounded NN controller]{
    \includegraphics[width=0.45\textwidth]{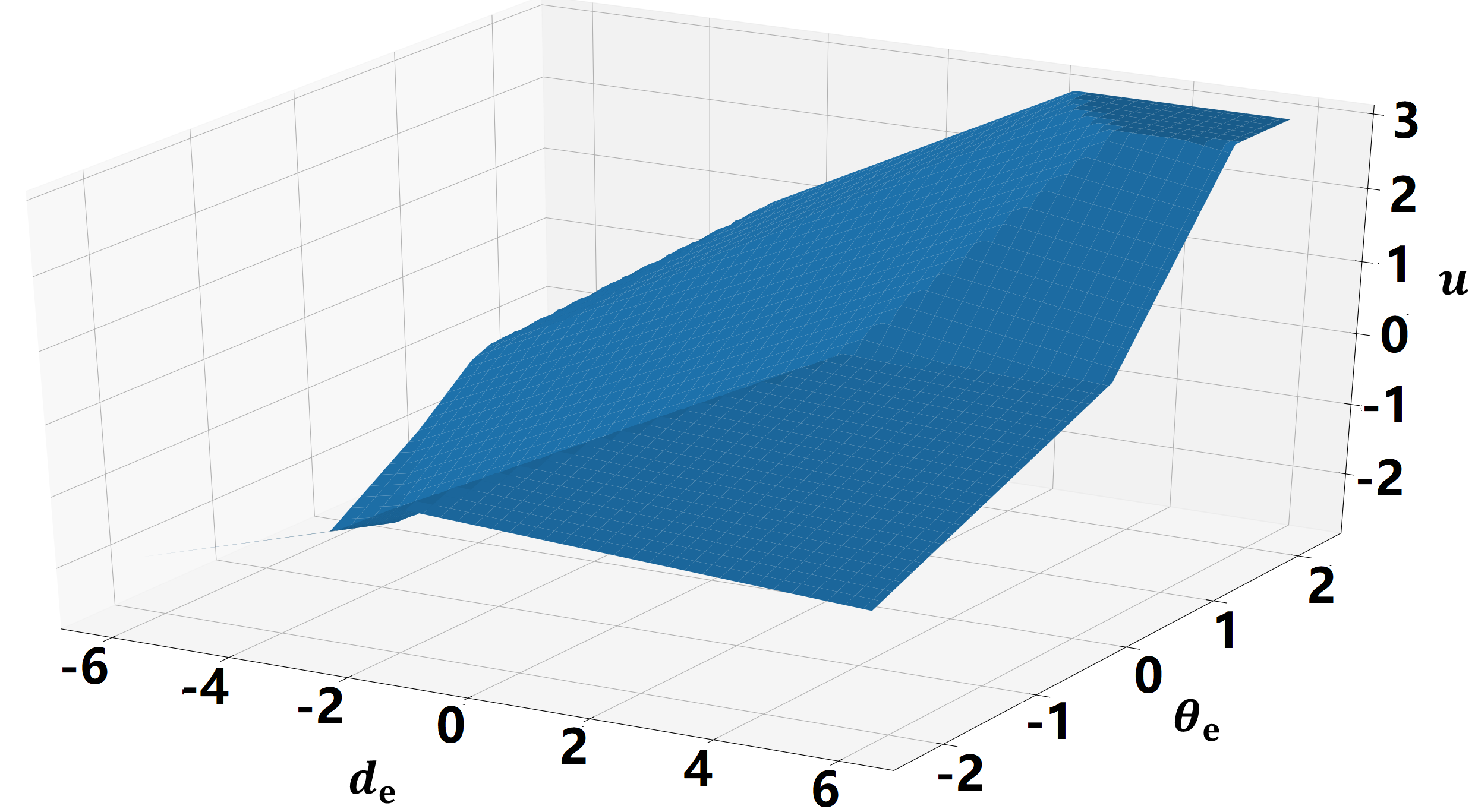}
    \label{fig:bounded-sub}}
    \caption{Plotting of surfaces of unbound or bounded NN controllers for Example~\ref{eg:dubins}  over $X_D$}\label{fig:bounded-ctrl}
\end{figure}

In practice, the control input $\mathbf u$ to system (\ref{eqn:control-system}) cannot take arbitrary values but are bounded within a compact set $U$.
Therefore it is necessary to consider how to synthesize bounded NN controllers for practical applications. Actually this can be achieved simply by replacing
the identity activation function in the output layer of $\mathcal N_{\rm c}$ (cf. Section~\ref{sec:nn-struc}) by any activation with bounded range, say hyperbolic tangent function.
For ease of formal verification, we adopt a piece-wise linear activation {\emph{Hardtanh}} for the output layer of $\mathcal N_{\rm c}$, that is,
\[a^{(L)}(x) = c \cdot \max\big(-1, \min(1, x)\big)\] 
with $c$ a positive constant, 
which restricts the output of $\mathcal N_{\rm c}$ to be within $[-c, c]$ for each dimension.
For Example~\ref{eg:dubins}, by choosing $c=3$ we learned a bounded NN controller as shown in Fig.~\ref{fig:bounded-sub}.
In our experiment, the Hardtanh activation can either be applied in the pre-training or fine-tuning process.


\section{Implementation and Experiments}\label{sec:implement}
Given a controlled CCDS $\Gamma=(\mathbf f, X_D, X_I, X_U)$ and generated training data set $S_D, S_I, S_U$,
in the most general form, the loss function we adopted for training safe NN controllers is:
\begin{equation}\label{eqn:cost-general}
  L(S_D, S_I, S_U) = c_1  \sum_{\mathbf x\in S_I} L_1(\mathbf x) + c_2 \sum_{\mathbf x\in S_U} L_2(\mathbf x) + 
  \sum_{\mathbf x\in S_D} \Big(c_3 L_3(\mathbf x) + c_4 L_4(\mathbf x)
  + c_5 L_5(\mathbf x)\Big) + c_6 L_6(\mathbf x_o)
\end{equation}
where $\mathbf x_o$ is the equilibrium point, $L_1,L_2,L_3,L_4,L_5,L_6$ are defined in (\ref{eqn:cost-123}), (\ref{eqn:cost-norm-lie}) and (\ref{eqn:loss-stablity}), 
$c_1,c_2,c_3$ are defined in (\ref{eqn:cost-full-data}), and $c_4,c_5,c_6$ are non-negative constant sub-loss weights. Thus there are
totally 6 sub-loss weights denoted by $\mathbf c = (c_1,c_2, \ldots, c_6)$ for short; besides, there are 8 tolerances in (\ref{eqn:cost-general})
denoted by $\boldsymbol{\varepsilon} = (\varepsilon_1, \varepsilon_2, \ldots, \varepsilon_8)$ for short. Our implementation and experiments are
conducted based on (\ref{eqn:cost-general}) and related notations.

\subsection{The Training Algorithm}
The main algorithm for training a safe NN controller is presented in Algorithm~\ref{alg:training},
\begin{algorithm}
\renewcommand{\algorithmicrequire}{\textbf{Input:}}
\renewcommand{\algorithmicensure}{\textbf{Output:}}
\caption{Safe NN-Controller Training Algorithm}
\label{alg:training}
\begin{algorithmic}[1]
\REQUIRE $\Gamma=(\mathbf f, X_D, X_I, X_U)$, $n_{\rm{restart}}$, 
$n_{\rm{epoch}}$, $n_{\rm{batch}}$, $l_r$, $\mathbf c$, $\pmb{\varepsilon}$;
\ENSURE $\mathcal N_{\rm c}$, $\mathcal N_{\rm b}$;
\STATE $\mathcal N_{\rm c}$, $\mathcal N_{\rm b}$ = \textsf{nn\_construct($\Gamma$)};
\STATE \textsf{data\_gen($\Gamma$)};
\FOR{$i=1$ to $n_{\rm{restart}}$}
\STATE \textsf{initialize($\mathcal N_{\rm c}$, $\mathcal N_{\rm b}$)};
\FOR{$j=1$ to $n_{\rm{epoch}}$}
\STATE $L_{\rm epoch} = 0$;
\FOR{$k=1$ to $n_{\rm{batch}}$}
\STATE $L_{\rm epoch}$ += \textsf{compute\_batch\_loss($\mathbf c$, $\pmb{\varepsilon}$)};
\STATE \textsf{update($\mathcal N_{\rm c}$, $\mathcal N_{\rm b}$, $l_r$)};
\ENDFOR
\IF{\textsf{decide\_success($L_{\rm epoch}$)}}
\RETURN{$\mathcal N_{\rm c}$, $\mathcal N_{\rm b}$;}
\ENDIF 
\ENDFOR
\ENDFOR
\end{algorithmic}
\end{algorithm}
which can be explained as follows:
\begin{itemize}
  \item $n_{\rm restart}, n_{\rm epoch}, n_{\rm batch}$ and $l_r$ are hyper-parameters for training (cf. Section~\ref{sec:training});
        in all our case studies, $n_{\rm restart}$ and $n_{\rm batch}$ are fixed at 5 and 4096 respectively;
  \item \textsf{nn\_construct()} in Line 1 is to construct the structure of $\mathcal N_{\rm c}$ and $\mathcal N_{\rm b}$ (cf. Section~\ref{sec:nn-struc});
        in all our case studies, $\mathcal N_{\rm c}$ has one hidden layer with 5 neurons, and $\mathcal N_{\rm b}$ has one hidden layer with 10 neurons;
  \item \textsf{data\_gen()} in Line 2 is to generate batches of training data (cf. Section~\ref{sec:data});
  \item \textsf{initialize()} in Line 4 is to initialize weights and biases of $\mathcal N_{\rm c}$ and $\mathcal N_{\rm b}$ by Gaussian distribution;
  \item \textsf{compute\_batch\_loss()} in Line 8 is to compute the loss value on each batch of data using the input $\mathbf c, \pmb{\varepsilon}$ \big(cf. Section~\ref{sec:loss} and (\ref{eqn:cost-general})\big);
  \item \textsf{update()} in Line 9 is to update $\mathcal N_{\rm c}$ and $\mathcal N_{\rm b}$ using \emph{gradient descent} with step-size $l_r$;
  \item \textsf{decide\_success()} in Line 11 is to decide the termination condition, which involves checking whether the epoch loss $L_{\rm epoch}$ reaches $0$.
\end{itemize}

We have implemented a prototype tool {\sf nncontroller}\footnote{Publically available at: \url{https://github.com/zhaohj2017/FAoC-tool}}
based on the {\sf{Pytorch}}\footnote{\url{https://pytorch.org/}} platform. Given a problem description and a set of user-specified parameters (cf. Algorithm~\ref{alg:training}), {\sf nncontroller} automatically
learns a safe NN controller with a NN barrier certificate, and generate script files as the input to {\sf iSAT3} for formal verification.
We have applied {\sf nncontroller} to a number of cases in the literature  \cite{Tuncali2018INVITEDRA,Deshmukh2019LearningDN,PLDI2019HZhu}. 
All experiments are performed on a laptop workstation running Ubuntu 18.04 with Intel
i7-8550u CPU and 32GB memory. The details of cases studies are presented in the following sub-section.

\subsection{Experiment Results}
\begin{figure}
  \centering
  \subfigure[The bicycle model]{
  \includegraphics[width=0.35\textwidth]{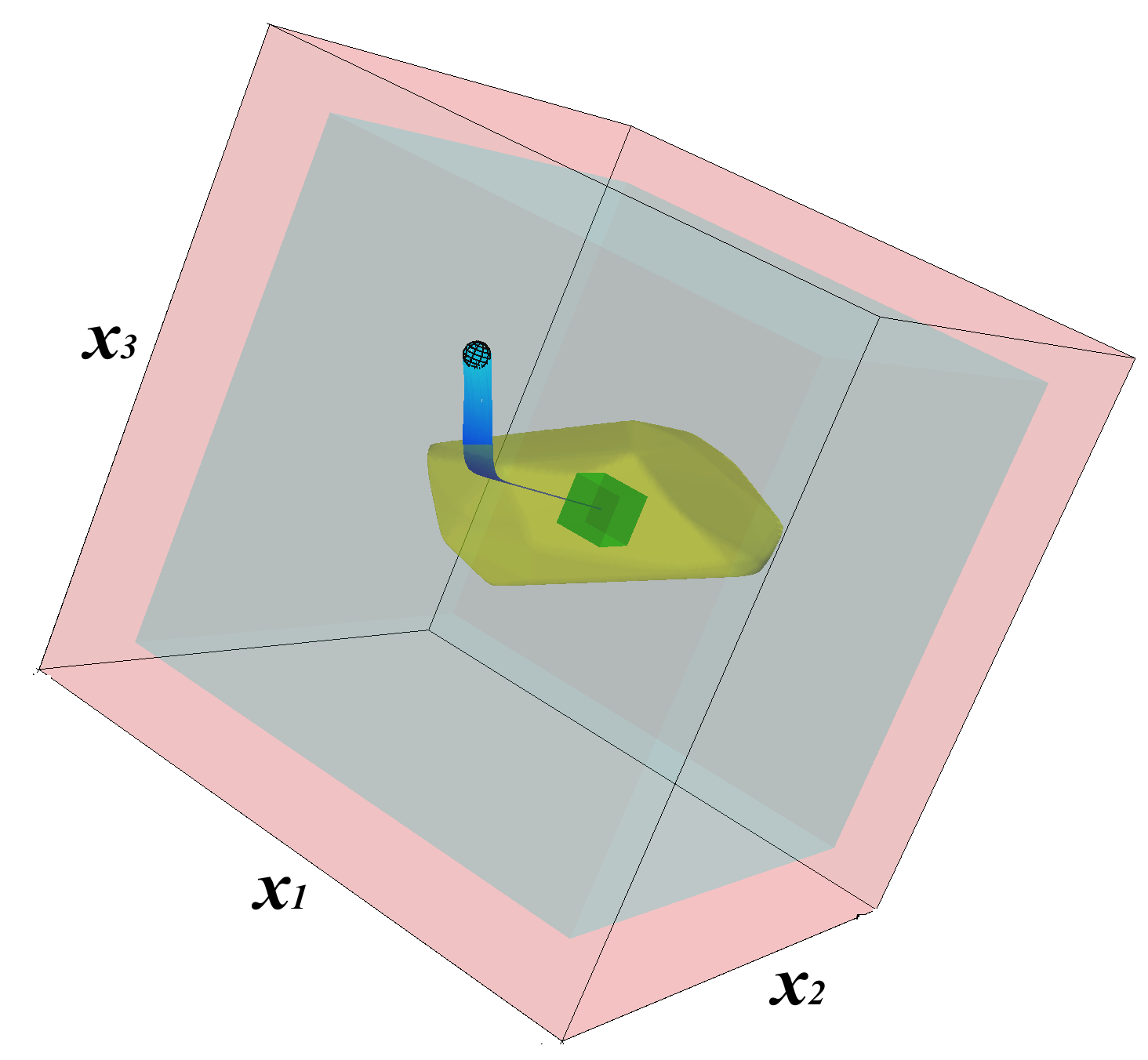}
  \label{fig:bicycle}}
  \qquad\qquad\qquad
  \subfigure[The academic 3D model]{
  \includegraphics[width=0.30\textwidth]{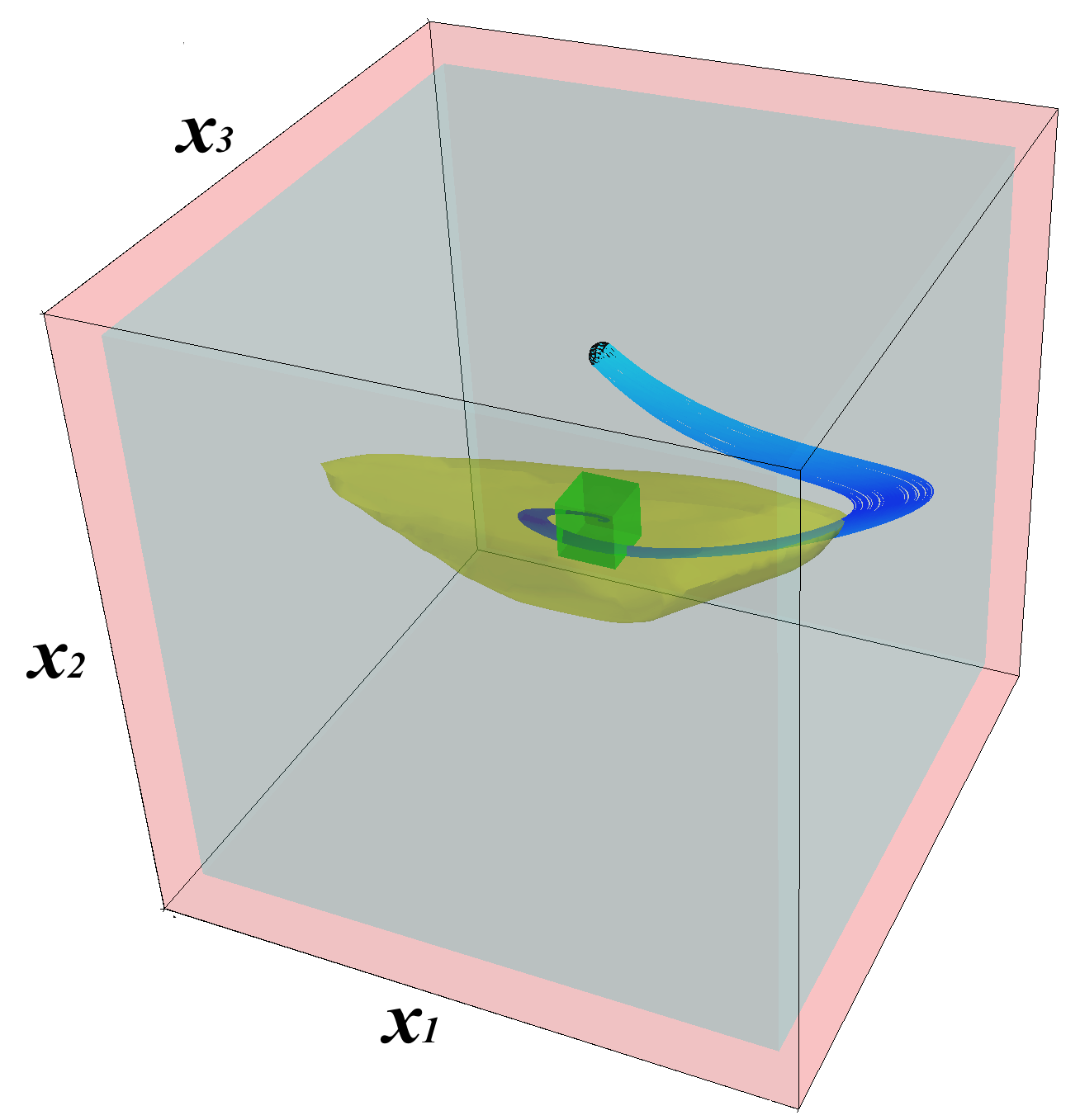}
  \label{fig:academic}}
  \caption{Learned and verified NN controllers and barriers for Example~\ref{eg:bicycle} and \ref{eg:academic}: for both cases, the innermost cube (green)
  represents the initial set, the outermost cube (pink) represents the system domain, and the space between the outermost and the middle cube (grey) is the unsafe region;
  the irregular surface (yellow) surrounding the innermost cube is the zero-level set of synthesized NN barrier; the curves (blue) approaching the origin are simulated system trajectories 
  }\label{fig:3d-barrier}
\end{figure}

In addition to the running example, we have synthesized and verified NN controllers using {\sf nncontroller} for the following cases.
\begin{example}[Inverted Pendulum \cite{PLDI2019HZhu}]\label{eg:inverted}
The controlled CCDS $\Gamma = (\mathbf f, X_D, X_I, X_U)$ is:
\begin{equation}\nonumber
\mathbf f:
\left[
    \begin{array}{l}
    \dot{\theta} \\ \dot{\omega}
    \end{array}
\right] =
\left[
    \begin{array}{c}
     \omega \\ \frac{g}{l}(\theta - \frac{\theta^3}{6}) + \frac{1}{ml^2}u
    \end{array}\right], 
\end{equation}
where $m=1$ and $l=1$ denote the pendulum mass and length respectively, $g=9.8$ is the gravitational acceleration, 
$u$ is the scalar control input maintaining the pendulum upright, and 
\begin{itemize}
\item $X_D$: $\{\theta, \omega)\in \mathbb R^2 | -\pi/2 \leq \theta \leq \pi / 2,\, -\pi / 2\leq \omega \leq \pi / 2\}$;
\item $X_I$: $\{\theta, \omega)\in \mathbb R^2 | -\pi/9 \leq \theta \leq \pi / 9,\, -\pi / 9\leq \omega \leq \pi / 9\}$;
\item $X_U$: the complement of $\{\theta, \omega)\in \mathbb R^2 | -\pi/6 \leq \theta \leq \pi / 6, \, -\pi / 6\leq \omega \leq \pi / 6\}$ in $X_D$.
\end{itemize}
\end{example}

\begin{example}[Duffing Oscillator \cite{PLDI2019HZhu}]\label{eg:duffing}
  The controlled CCDS $\Gamma = (\mathbf f, X_D, X_I, X_U)$ is:
  \begin{equation}\nonumber
  \mathbf f:
  \left[
      \begin{array}{l}
      \dot{x} \\ \dot{y}
      \end{array}
  \right] =
  \left[
      \begin{array}{c}
       y\\ -0.6y - x - x^3 + u
      \end{array}\right], 
  \end{equation}
where $u$ is the scalar control input that regulates the system's trajectories to $(0, 0)$, and
  \begin{itemize}
  \item $X_D$: $\{x, y)\in \mathbb R^2 | -6\leq x\leq 6,\, -6\leq y \leq 6 \}$;
  \item $X_I$: $\{x, y)\in \mathbb R^2 | -2.5\leq x\leq 2.5,\, -2\leq y \leq 2 \}$;
  \item $X_U$: the complement of $\{x, y)\in \mathbb R^2 | -5\leq x\leq 5,\, -5\leq y \leq 5 \}$ in $X_D$.
  \end{itemize}
\end{example}

\begin{example}[Bicycle Steering \cite{Deshmukh2019LearningDN}]\label{eg:bicycle}
The control objective is to balance a bicycle.
The states of the bicycle are $(x_1, x_2, x_3)$ which denote the tilt angle, the angular velocity of tilt, 
and the handle bar angle with body respectively. The controlled CCDS $\Gamma = (\mathbf f, X_D, X_I, X_U)$ is:
\begin{equation}\nonumber
  \mathbf f:
  \left[
      \begin{array}{l}
      \dot{x_1} \\ \dot{x_2} \\ \dot{x_3}
      \end{array}
  \right] =
  \left[
      \begin{array}{c}
       x_2 \\ \frac{ml}{J} (g\sin{x_1} + \frac{v^2}{b}\cos{x_1}\tan{x_3} ) \\ 0
      \end{array}\right]
      +
  \left[
    \begin{array}{c}
    0\\ \frac{amlv}{Jb} \cdot \frac{\cos{x_1}}{{\cos{\!^2x_3}}}   \\1
    \end{array}
  \right] u\,,
  \end{equation}
  where $u$ is the scalar control input, $m=20$ is the mass, $l=1$ is the height, $b=1$ is the wheel base, $J=\frac{mb^2}{3}$
  is the moment of inertia, $v=10$ is the velocity, $g=10$ is the acceleration of gravity, $a = 0.5$, and 
  \begin{itemize}
  \item $X_D$: $\{(x_1,x_2,x_3)\in \mathbb R^3 | -2.2\leq x_1\leq 2.2, -2.2\leq x_2\leq 2.2, -2.2\leq x_3\leq 2.2\}$;
  \item $X_I$: $\{(x_1,x_2,x_3)\in \mathbb R^3 | -0.2\leq x_1\leq 0.2, -0.2\leq x_2\leq 0.2, -0.2\leq x_3\leq 0.2\}$;
  \item $X_U$: the complement of $\{(x_1,x_2,x_3)\in \mathbb R^3 | -2\leq x_1\leq 2, -2\leq x_2\leq 2, -2\leq x_3\leq 2\}$ in $X_D$.
  \end{itemize}
  By introducing $\tilde u$ such that $u = \tilde u \cos{\!^2x_3} - 20 \cos{x_3}\sin{x_3}$, the original $\mathbf f$ is transformed equivalently into
  \begin{equation}\nonumber
    \mathbf {\tilde f}:
    \left[
        \begin{array}{l}
        \dot{x_1} \\ \dot{x_2} \\ \dot{x_3}
        \end{array}
    \right] =
    \left[
        \begin{array}{c}
         x_2 \\ 30\sin{x_1} + 15 \tilde u\cos{x_1}   \\ \tilde u \cos{\!^2x_3} - 20 \cos{x_3}\sin{x_3}
        \end{array}\right].
    \end{equation}
\end{example}
A NN controller representing $\tilde u$ was learned and verified for the transformed system 
$({\mathbf {\tilde f}}, X_D, X_I, X_U)$ (cf. Fig.~\ref{fig:bicycle}).

\begin{example}[Academic 3D \cite{Deshmukh2019LearningDN}]\label{eg:academic}
The controlled CCDS $\Gamma = (\mathbf f, X_D, X_I, X_U)$ is:
\begin{equation}\nonumber
\mathbf f:
\left[
    \begin{array}{l}
    \dot{x_1} \\ \dot{x_2} \\ \dot{x_3}
    \end{array}
\right] =
\left[
    \begin{array}{c}
     x_3 + 8 x_2 \\ -x_2 + x_3 \\ -x_3 - x_1^2
    \end{array}\right]
    +
\left[
  \begin{array}{c}
  0\\0\\1
  \end{array}
\right] u\,,
\quad \textrm{where } u \textrm{ is the scalar control input}
\end{equation}
\begin{itemize}
\item $X_D$: $\{(x_1,x_2,x_3)\in \mathbb R^3 | -2.2\leq x_1\leq 2.2, -2.2\leq x_2\leq 2.2, -2.2\leq x_3\leq 2.2\}$;
\item $X_I$: $\{(x_1,x_2,x_3)\in \mathbb R^3 | -0.2\leq x_1\leq 0.2, -0.2\leq x_2\leq 0.2, -0.2\leq x_3\leq 0.2\}$;
\item $X_U$: the complement of $\{(x_1,x_2,x_3)\in \mathbb R^3 | -2\leq x_1\leq 2, -2\leq x_2\leq 2, -2\leq x_3\leq 2\}$ in $X_D$.
\end{itemize}
A NN controller was successfully learned and verified for $\Gamma$ (cf. Fig.~\ref{fig:academic}).
\end{example}

\begin{table}
  \caption{Key parameters for pre-training and fine-tuning by {\sf nncontroller} (cf. Algorithm~\ref{alg:training} and Remark~\ref{rmk:table1})}\label{tbl:para}
  \begin{tabular}{@{}lllllll@{}}
   \hline
   E.g.                 & $n_{\rm e}$        & $l_r$                & \hspace{.8cm}$\mathbf c$     & \hspace{1.5cm}$\pmb \varepsilon$                     & \hspace{.8cm}${\mathbf c}^{\rm v}$   & \hspace{1.5cm}${\pmb \varepsilon}^{\rm v}$  \\
   \hline
     \ref{eg:dubins}    & 100                & 0.1                  & $(1,1,1,0,0,0)$              & $(0,0,0,0.01,\cdot,\cdot,\cdot,\cdot)$               & $(1,1,1,0,0,0)$                      & $(0,0,0.01,0.01,\cdot,\cdot,\cdot,\cdot)$  \\
   \hline
     \ref{eg:inverted}  & 100                & 0.1                  & $(1,1,1,0,0,0)$              & $(0,0,0,0.01,\cdot,\cdot,\cdot,\cdot)$               & $(1,1,1,0,0,0)$                      & $(0.01,0,0.02,0.01,\cdot,\cdot,\cdot,\cdot)$  \\
   \hline
     \ref{eg:duffing}   & 100                & 0.01$\leadsto$0.1    & $(1,1,1,0,0,0)$              & $(0,0,0,0.05,\cdot,\cdot,\cdot,\cdot)$               & $(1,1,1,0,0,0)$                      & $(0,0,0,0.05,\cdot,\cdot,\cdot,\cdot)$  \\
   \hline
     \ref{eg:bicycle}   & 200                & 0.01$\leadsto$0.2    & $(1,1,0.1,0.1,0,0)$          & $(0,0,0,0.02,0,\cdot,\cdot,\cdot)$                   & $(1,1,0.1,0.1,0.01,0.01)$            & $\scriptstyle (0,0,0.35,0.02,0.35,0.1,0.1,0.01)$  \\
   \hline
     \ref{eg:academic}  & 200                & 0.01$\leadsto$0.2    & $(1,1,0.1,0.1,0,0)$          & $(0,0,0,0.02,0,\cdot,\cdot,\cdot)$                   & $(1,1,0.1,0.1,0.01,0.01)$            & $\scriptstyle (0.01,0.01,0.15,0.02,0.1,0.1,0.2,0.01)$  \\
   \hline
  \end{tabular}
\end{table}
\begin{table}
  \caption{Time costs of synthesis and verification by {\sf nncontroller} and {\sf iSAT3} (cf. Remark~\ref{rmk:table2})}\label{tbl:time}
  \begin{tabular}{@{}lllllllllllll@{}}
   \hline
   \multirow{2}{1cm}{E.g.}& \multicolumn{2}{c}{run 1}         & \multicolumn{2}{c}{run 2}             & \multicolumn{2}{c}{run 3}         & \multicolumn{2}{c}{run 4}        & \multicolumn{2}{c}{run 5}      & \multirow{2}{1.5cm}{learning avg. cost}  & \multirow{2}{2cm}{verification cost} \\
                           \cline{2-11}
                           & time \hspace{.3cm}& $n_{\rm r}\hspace{.3cm}$                  & time \hspace{.3cm}& $n_{\rm r}\hspace{.3cm}$                      & time \hspace{.3cm}& $n_{\rm r}\hspace{.3cm}$                  & time \hspace{.3cm}& $n_{\rm r}\hspace{.3cm}$                 & time \hspace{.3cm}& $n_{\rm r}\hspace{.3cm}$              &                                    &                 \\
   \hline
     \ref{eg:dubins}       & 21.11 & 0                          & 15.04 & 0                              & 14.98 & 0                          & 65.25 & 0                         & 15.37  & 0                     & 26.35                            & 8.27  \\
   \hline
     \ref{eg:inverted}     & 478.29 & 1                         & 168.75 & 0                             & 292.96 & 0                         & 111.55 & 0                        & 43.89 & 0                      & 219.09                           & 15.24  \\
   \hline
     \ref{eg:duffing}      & 60.59 & 0                          & 72.47 & 0                              & 64.64 & 0                          & 48.08 & 0                         & 851.49 & 1                     & 219.45                           & 4.71  \\
   \hline
     \ref{eg:bicycle}      & 752.63 & 1                         & 1528.07 & 2                            & 499.83 & 0                         & 122.64 & 0                        & 924.41 & 1                     & 765.52                           & 1344.50  \\
   \hline
     \ref{eg:academic}     & 240.94 & 0                         & 301.22 & 0                             & 2522.14 & 3                        & 1001.66 & 1                       & 390.25 & 0                     & 891.24                           & 6070.83  \\
   \hline
  \end{tabular}
\end{table}

The key parameters used by {\sf nncontroller} for our experiments are summarized in Table~\ref{tbl:para}, and the time costs of synthesis
and verification by {\sf nncontroller} and {\sf iSAT3} are summarized in Table~\ref{tbl:time}. 
\begin{remark}\label{rmk:table1}
 In Table~\ref{tbl:para}, $n_{\rm e}$ is a shorthand for $n_{\rm epoch}$, $\cdot$ means the corresponding parameter is not applicable, 
$\leadsto$ means we adopt a self-adaptive learning rate scheduling strategy, and the superscript $\rm v$ means that the weight coefficients  
${\mathbf c}^{\rm v}$ and 
parameters ${\pmb \varepsilon}^{\rm v}$ are for the fine-tuned controllers, which are formally verified.
\end{remark}
\begin{remark}\label{rmk:table2}
  In Table~\ref{tbl:time}, all time costs are measured in seconds; the time cost of NN controller training is not deterministic since the NN models are initialized randomly 
  and the batches of training data are shuffled during the training process, and therefore we record the time costs of 5 separate runs of the
  training algorithm and compute the averaged cost; $n_{\rm r}$ denotes how many times we restart the algorithm when no NN controller is learned within the specified number
  of training epochs, i.e. $n_{\rm epoch}$; the last column corresponds to time costs of formal verification for the NN controllers and barriers obtained with the ${\mathbf c}^{\rm v}$ 
  and ${\pmb \varepsilon}^{\rm v}$ parameters in Table~\ref{tbl:para} for each case.
\end{remark}
\begin{remark} Comparison of time costs of our experiment with related work such as \cite{Deshmukh2019LearningDN,PLDI2019HZhu} is not straightforward
  since we train two NNs simultaneously, while \cite{Deshmukh2019LearningDN} requires user-provided barrier functions and \cite{PLDI2019HZhu} requires
  pre-trained NN controllers as their inputs. However, considering the number of layers and neurons (we use one hidden layer with 5 neurons and ReLU activations for $\mathcal N_{\rm c}$ uniformly), it can be asserted that our synthesized NN controllers have much simpler structure
  than \cite{Deshmukh2019LearningDN,PLDI2019HZhu}.
\end{remark}

\section{Conclusion}\label{sec:conclude}
We have proposed a new approach to synthesize neural network controllers for nonlinear continuous dynamical systems with
control against safety properties. Our approach features in verification-in-the-loop synthesis: we simultaneously train the controller and its certificate, which we use barrier functions, represented by an NN as well. We have provided a prototype tool {\sf nncontroller} with a number of case studies. 
The experiment results have confirmed the feasibility and efficacy of our approach.

Future work includes experimenting on different sampling and training strategies to reduce the data set size and to improve the training efficiency, as well different verification methods/tools other than interval SMT solvers. We anticipate that these would potentially further improve the scalability of our approach. We also plan to extend our approach to other properties such as reachability coupled with cost/reward based optimality as what has been done in optimal control.



\paragraph{Acknowledgements.} 
We thank Prof. Jyotirmoy V. Deshmukh for explanations on the bicycle model of Example~\ref{eg:bicycle}. 

\small{H. Zhao was supported partially by the National Natural
Science Foundation of China (No. 61702425, 61972385); 
X. Zeng was supported partially by the National Natural
Science Foundation of China (No. 61902325), and 
``Fundamental Research Funds for the Central Universities" (SWU117058);
T. Chen is partially supported by NSFC grant (No. 61872340), and Guangdong Science and Technology Department grant (No. 2018B010107004), the Overseas Grant of the State Key Laboratory of Novel Software Technology (No. KFKT2018A16), the Natural Science Foundation of Guangdong Province of China (No. 2019A1515011689);
Z. Liu was supported partially by the National Natural
Science Foundation of China (No. 61672435, 61732019, 61811530327), and Capacity Development Grant of Southwest University (SWU116007);
J. Woodcock was partially supported by the research grant from Southwest University.}


\begin{thebibliography}{10}
  \providecommand{\url}[1]{\texttt{#1}}
  \providecommand{\urlprefix}{URL }
  \providecommand{\doi}[1]{https://doi.org/#1}
  
  \bibitem{ahmadi2019safe}
  Ahmadi, M., Singletary, A., Burdick, J.W., Ames, A.D.: Safe policy synthesis in
    multi-agent {POMDP}s via discrete-time barrier functions. In: 2019 IEEE 58th
    Conference on Decision and Control (CDC). pp. 4797--4803. IEEE (2019)
  
  \bibitem{CBF19}
  {Ames}, A.D., {Coogan}, S., {Egerstedt}, M., {Notomista}, G., {Sreenath}, K.,
    {Tabuada}, P.: Control barrier functions: Theory and applications. In: 2019
    18th European Control Conference (ECC). pp. 3420--3431 (2019)
  
  \bibitem{Berkenkamp17-nips}
  Berkenkamp, F., Turchetta, M., Schoellig, A.P., Krause, A.: Safe model-based
    reinforcement learning with stability guarantees. In: Proceedings of the 31st
    International Conference on Neural Information Processing Systems. p.
    908–919. NIPS’17, Curran Associates Inc., Red Hook, NY, USA (2017)
  
  \bibitem{GAO-NIPS2019}
  Chang, Y.C., Roohi, N., Gao, S.: Neural {L}yapunov control. In: Advances in
    Neural Information Processing Systems 32, pp. 3245--3254. Curran Associates,
    Inc. (2019)
  
  \bibitem{End-to-end19}
  Cheng, R., Orosz, G., Murray, R.M., Burdick, J.W.: End-to-end safe
    reinforcement learning through barrier functions for safety-critical
    continuous control tasks. In: The Thirty-Third {AAAI} Conference on
    Artificial Intelligence, {AAAI} 2019, Honolulu, Hawaii, USA, January 27 -
    February 1, 2019. pp. 3387--3395. {AAAI} Press (2019)
  
  \bibitem{choi2020reinforcement}
  Choi, J., Castañeda, F., Tomlin, C.J., Sreenath, K.: Reinforcement learning
    for safety-critical control under model uncertainty, using control {L}yapunov
    functions and control barrier functions (2020)
  
  \bibitem{dai2013barrier}
  Dai, L., Gan, T., Xia, B., Zhan, N.: Barrier certificates revisited. Journal of
    Symbolic Computation  \textbf{80},  62--86 (2017)
  
  \bibitem{Deshmukh2019LearningDN}
  Deshmukh, J.V., Kapinski, J., Yamaguchi, T., Prokhorov, D.: Learning deep
    neural network controllers for dynamical systems with safety guarantees:
    Invited paper. In: 2019 IEEE/ACM International Conference on Computer-Aided
    Design (ICCAD). pp.~1--7 (2019)
  
  \bibitem{SESHIA19}
  Dreossi, T., Fremont, D.J., Ghosh, S., Kim, E., Ravanbakhsh, H.,
    Vazquez-Chanlatte, M., Seshia, S.A.: {VerifAI}: A toolkit for the formal
    design and analysis of artificial intelligence-based systems. In: Computer
    Aided Verification. pp. 432--442. Springer International Publishing (2019)
  
  \bibitem{OpenAI-DRL}
  Duan, Y., Chen, X., Houthooft, R., Schulman, J., Abbeel, P.: Benchmarking deep
    reinforcement learning for continuous control. In: Proceedings of the 33nd
    International Conference on Machine Learning, {ICML} 2016, New York City, NY,
    USA, June 19-24, 2016. {JMLR} Workshop and Conference Proceedings, vol.~48,
    pp. 1329--1338. JMLR.org (2016)
  
  \bibitem{DBLPSouradeep19HSCC}
  Dutta, S., Chen, X., Sankaranarayanan, S.: Reachability analysis for neural
    feedback systems using regressive polynomial rule inference. In: Proceedings
    of the 22nd {ACM} International Conference on Hybrid Systems: Computation and
    Control, {HSCC}. pp. 157--168 (2019)
  
  \bibitem{DUTTA2018151}
  Dutta, S., Jha, S., Sankaranarayanan, S., Tiwari, A.: Learning and verification
    of feedback control systems using feedforward neural networks.
    IFAC-PapersOnLine  \textbf{51}(16),  151 -- 156 (2018), 6th IFAC Conference
    on Analysis and Design of Hybrid Systems ADHS 2018
  
  \bibitem{Dutta2018}
  Dutta, S., Jha, S., Sankaranarayanan, S., Tiwari, A.: Output range analysis for
    deep feedforward neural networks. In: NASA Formal Methods. pp. 121--138.
    Springer International Publishing (2018)
  
  \bibitem{FultonP18}
  Fulton, N., Platzer, A.: Safe reinforcement learning via formal methods: Toward
    safe control through proof and learning. In: Proceedings of the Thirty-Second
    {AAAI} Conference on Artificial Intelligence, (AAAI-18), New Orleans,
    Louisiana, USA, February 2-7, 2018. pp. 6485--6492. {AAAI} Press (2018)
  
  \bibitem{dl-ian}
  Goodfellow, I., Bengio, Y., Courville, A.: Deep Learning. The MIT Press (2016)
  
  \bibitem{linear-system-2}
  Hespanha, J.P.: Linear Systems Theory. Princeton University Press, second edn.
    (2018)
  
  \bibitem{PappasHSCC20}
  Ivanov, R., Carpenter, T.J., Weimer, J., Alur, R., Pappas, G.J., Lee, I.: Case
    study: verifying the safety of an autonomous racing car with a neural network
    controller. In: {HSCC} '20: 23rd {ACM} International Conference on Hybrid
    Systems: Computation and Control, Sydney, New South Wales, Australia, April
    21-24, 2020. pp. 28:1--28:7. {ACM} (2020)
  
  \bibitem{IvanovWAPL19}
  Ivanov, R., Weimer, J., Alur, R., Pappas, G.J., Lee, I.: Verisig: verifying
    safety properties of hybrid systems with neural network controllers. In:
    Proceedings of the 22nd {ACM} International Conference on Hybrid Systems:
    Computation and Control, {HSCC} 2019. pp. 169--178 (2019)
  
  \bibitem{jordan2020exactly}
  Jordan, M., Dimakis, A.G.: Exactly computing the local {L}ipschitz constant of
    {ReLU} networks (2020)
  
  \bibitem{katz2017reluplex}
  Katz, G., Barrett, C., Dill, D.L., Julian, K., Kochenderfer, M.J.: Reluplex: An
    efficient smt solver for verifying deep neural networks. In: International
    Conference on Computer Aided Verification. pp. 97--117. Springer (2017)
  
  \bibitem{kong2013exponential}
  Kong, H., He, F., Song, X., Hung, W.N., Gu, M.: Exponential-condition-based
    barrier certificate generation for safety verification of hybrid systems. In:
    Proceedings of the 25th International Conference on Computer Aided
    Verification (CAV). pp. 242--257. Springer (2013)
  
  \bibitem{LESHNO1993861}
  Leshno, M., Lin, V.Y., Pinkus, A., Schocken, S.: Multilayer feedforward
    networks with a nonpolynomial activation function can approximate any
    function. Neural Networks  \textbf{6}(6),  861 -- 867 (1993)
  
  \bibitem{Chenlq19SAS}
  Li, J., Liu, J., Yang, P., Chen, L., Huang, X., Zhang, L.: Analyzing deep
    neural networks with symbolic propagation: Towards higher precision and
    faster verification. In: Static Analysis. pp. 296--319. Springer
    International Publishing (2019)
  
  \bibitem{DeepM-DRL}
  Lillicrap, T.P., Hunt, J.J., Pritzel, A., Heess, N., Erez, T., Tassa, Y.,
    Silver, D., Wierstra, D.: Continuous control with deep reinforcement
    learning. In: 4th International Conference on Learning Representations,
    {ICLR} 2016, San Juan, Puerto Rico, May 2-4, 2016, Conference Track
    Proceedings (2016)
  
  \bibitem{manek2020learning}
  Manek, G., Kolter, J.Z.: Learning stable deep dynamics models (2020)
  
  \bibitem{mittal2020neural}
  Mittal, M., Gallieri, M., Quaglino, A., Salehian, S.S.M., Koutník, J.: Neural
    {L}yapunov model predictive control (2020)
  
  \bibitem{NNforControl}
  Poznyak, A., EN, S., Yu, W.: Differential Neural Networks for Robust Nonlinear
    Control. World Scientific (2001)
  
  \bibitem{prajna2007safety}
  Prajna, S., Jadbabaie, A., Pappas, G.J.: A framework for worst-case and
    stochastic safety verification using barrier certificates. IEEE Transactions
    on Automatic Control  \textbf{52}(8),  1415--1429 (2007)
  
  \bibitem{Pulina2010CAV}
  Pulina, L., Tacchella, A.: An abstraction-refinement approach to verification
    of artificial neural networks. In: Computer Aided Verification. pp. 243--257
    (2010)
  
  \bibitem{Ratschan18-tac}
  {Ratschan}, S.: Converse theorems for safety and barrier certificates. IEEE
    Transactions on Automatic Control  \textbf{63}(8),  2628--2632 (2018)
  
  \bibitem{lya-sriram-19}
  Ravanbakhsh, H., Sankaranarayanan, S.: Learning control {L}yapunov functions
    from counterexamples and demonstrations. Autonomous Robots  \textbf{43}(2),
    275--307 (2019)
  
  \bibitem{saferl-openai}
  Ray, A., Achiam, J., Amodei, D.: Benchmarking safe exploration in deep
    reinforcement learning, \url{https://cdn.openai.com/safexp-short.pdf}
  
  \bibitem{lya-nn-2018}
  Richards, S.M., Berkenkamp, F., Krause, A.: The {L}yapunov neural network:
    Adaptive stability certification for safe learning of dynamic systems. CoRR
    \textbf{abs/1808.00924} (2018), \url{http://arxiv.org/abs/1808.00924}
  
  \bibitem{sloth2012compositional}
  Sloth, C., Pappas, G.J., Wisniewski, R.: Compositional safety analysis using
    barrier certificates. In: Proc. of the Hybrid Systems: Computation and
    Control (HSCC). pp. 15--24. ACM (2012)
  
  \bibitem{Sogokon2018FM}
  Sogokon, A., Ghorbal, K., Tan, Y.K., Platzer, A.: Vector barrier certificates
    and comparison systems. In: Formal Methods. pp. 418--437 (2018)
  
  \bibitem{SunKS19}
  Sun, X., Khedr, H., Shoukry, Y.: Formal verification of neural network
    controlled autonomous systems. In: Proceedings of the 22nd {ACM}
    International Conference on Hybrid Systems: Computation and Control, {HSCC}
    2019. pp. 147--156 (2019)
  
  \bibitem{Episodic19}
  {Taylor}, A.J., {Dorobantu}, V.D., {Le}, H.M., {Yue}, Y., {Ames}, A.D.:
    Episodic learning with control {L}yapunov functions for uncertain robotic
    systems. In: 2019 IEEE/RSJ International Conference on Intelligent Robots and
    Systems (IROS). pp. 6878--6884 (2019)
  
  \bibitem{taylor2019learning}
  Taylor, A., Singletary, A., Yue, Y., Ames, A.: Learning for safety-critical
    control with control barrier functions (2019)
  
  \bibitem{NNV-CAV20}
  Tran, H.D., Yang, X., Lopez, D.M., Musau, P., Nguyen, L.V., Xiang, W., Bak, S.,
    Johnson, T.T.: {NNV}: The neural network verification tool for deep neural
    networks and learning-enabled cyber-physical systems. In: 32nd International
    Conference on Computer-Aided Verification (CAV) (2020)
  
  \bibitem{Tuncali2018INVITEDRA}
  Tuncali, C.E., Kapinski, J., Ito, H., Deshmukh, J.V.: Invited: Reasoning about
    safety of learning-enabled components in autonomous cyber-physical systems.
    In: 2018 55th ACM/ESDA/IEEE Design Automation Conference (DAC). pp.~1--6
    (2018)
  
  \bibitem{WengZCSHDBD18}
  Weng, T., Zhang, H., Chen, H., Song, Z., Hsieh, C., Daniel, L., Boning, D.S.,
    Dhillon, I.S.: Towards fast computation of certified robustness for relu
    networks. In: Proceedings of the 35th International Conference on Machine
    Learning, {ICML} 2018. pp. 5273--5282 (2018)
  
  \bibitem{WeimingXiang2017}
  Xiang, W., Tran, H., Johnson, T.T.: Output reachable set estimation and
    verification for multi-layer neural networks. CoRR  \textbf{abs/1708.03322}
    (2017)
  
  \bibitem{yaghoubi2020training}
  Yaghoubi, S., Fainekos, G., Sankaranarayanan, S.: Training neural network
    controllers using control barrier functions in the presence of disturbances
    (2020)
  
  \bibitem{ZhaoZC020}
  Zhao, H., Zeng, X., Chen, T., Liu, Z.: Synthesizing barrier certificates using
    neural networks. In: {HSCC} '20. pp. 25:1--25:11. {ACM} (2020)
  
  \bibitem{PLDI2019HZhu}
  Zhu, H., Xiong, Z., Magill, S., Jagannathan, S.: An inductive synthesis
    framework for verifiable reinforcement learning. In: Proceedings of the 40th
    ACM SIGPLAN Conference on Programming Language Design and Implementation. pp.
    686--701. PLDI 2019, Association for Computing Machinery, New York, NY, USA
    (2019)
  
  \end{thebibliography}
\end{document}